\documentclass[preprint,12pt]{elsarticle}

\usepackage{graphicx}
\usepackage[all]{xy}
\usepackage{amsmath}
\usepackage{amssymb}
\usepackage{epstopdf}
\usepackage[colorlinks=true, pdfstartview=FitV, linkcolor=blue, citecolor=red, urlcolor=magenta]{hyperref}
\usepackage{latexsym}
\usepackage{amsfonts}
\usepackage{subeqnarray}

\newcommand{\be}{\begin{equation}}
\newcommand{\ee}{\end{equation}}
\newcommand{\ben}{\begin{eqnarray}}
\newcommand{\een}{\end{eqnarray}}
\newcommand{\bsen}{\begin{subeqnarray}}
\newcommand{\esen}{\end{subeqnarray}}

\newcommand{\bb}{\bibitem}
\newcommand{\pa}{\partial}
\newcommand{\bea}{\begin{eqnarray}}
\newcommand{\eea}{\end{eqnarray}}

\journal{Annals of Physics}

\begin{document}

\begin{frontmatter}



\title{Soliton solutions in two-dimensional Lorentz-violating higher derivative scalar theory}


\author[addr1]{E. Passos}
\ead{passos@df.ufcg.edu.br}
\author[addr3]{C. A. G. Almeida}
\ead{cviniro@gmail.com}
\author[addr1,addr4]{F. A. Brito}
\ead{fabrito@df.ufcg.edu.br}
\author[addr3,addr4]{R. Menezes}
\ead{betomenezes@gmail.com}
\author[addr5]{J. C. Mota-Silva}
\ead{julio.silva@ifma.edu.br}
\author[addr1]{J. R. L. Santos}
\ead{joaorafael@df.ufcg.edu.br}

\address[addr1]{Departamento de F\'\i sica, Universidade Federal de Campina Grande, Caixa Postal 10071,
58109-970  Campina Grande, Para\'\i ba, Brazil}
\address[addr3]{Departamento de Ci\^encias Exatas, Universidade Federal da Para\'\i  ba, 58297-000, Rio Tinto, Para\'\i ba, Brazil.}
\address[addr4]{Departamento de F\'\i sica, Universidade Federal da Para\'iba, Caixa Postal 5008, 58051-970, Jo\~ ao Pessoa, Para\' iba, Brazil.}
\address[addr5]{
Instituto Federal do Maranh\~ao, Campus Porto Franco, R. Custodio Barbosa, 65970-000, Porto Franco, Maranh\~ao, Brazil. }

\begin{abstract}
This paper shows a new approach to obtain analytical topological defects of a $2D$ Myers-Pospelov Lagrangian for two scalar fields. Such a Lagrangian presents higher-order kinetic terms, which lead us to equations of motion which are non-trivial to be integrated. Here we describe three possible scenarios for the equations of motion, named by timelike, spacelike and lightlike respectively. We started our investigation with a kink-like travelling wave {\it Ansatz} for the free theory, which led us to constraints for the dispersion relations of each scenario. We also introduced a procedure to obtain analytical solutions for the general theory in the three mentioned scenarios. We exemplified the procedure and discussed the behavior of the defect solutions carefully. It is remarkable that the methodology presented in this study led to analytical models, despite the complexity of the equations of motion derived from the 2D Myers-Pospelov Lagrangian. The methodology here tailored can be applied to several Lagrangians with higher-order derivative terms. 

\end{abstract}


\end{frontmatter}


\section{Introduction}

The perspective of a new physics at Planck energy scale ($M_{ P}\sim 10^{19}$ GeV), may lead us to scenarios described by Lorentz violating theories. In such level of energy both Lorentz and Charge-Parity-Time (CPT) symmetries are expected to be broken. One of the main works which describes these scenarios is Myers and Pospelov seminal paper \cite{myers}, where it was considered cubic modifications for the dispersion relations related with  mass operators of dimension-five along with a constant four-vector $n_{\mu}$ interacting with scalar, fermion and photon fields. In order to explore the effects of these new theories, the authors introduced an electromagnetic Lagrangian by a modified kinetic term of the following form:
$
{\cal L}_{MP}= g\,n^{\mu}\,F_{\mu\nu}\big(n\,\cdot\,\partial\big)n_{\alpha}\widetilde{F}^{\alpha\nu}\,,
$
where $F_{\mu\nu}=\pa_{\mu}A_{\nu} - \pa_{\nu}A_{\mu}$ is the electromagnetic field strength tensor and 
$\widetilde{F}^{\alpha\beta}=(1/2)\,\varepsilon^{\alpha\beta\lambda\rho}\,F_{\lambda\rho}$ its dual $U(1)$ gauge field tensor. Moreover, $g=\xi/M_{P}$, with $\xi$ being a dimensionless parameter suppressed by the Planck scale $M_{P}$. Notice also that by using the Bianchi  identity and the gauge conditions: $\pa\cdot A=0$, the previous Lagrangian reduces to ${\cal L}_{MP} = g\, \varepsilon^{\mu\nu\lambda\rho}n_{\mu} (n\cdot\pa)^{2} (\pa_{\nu}A_{\lambda})A_{\rho}$. The mentioned paper conjectured that the previous field theory provides a framework where one can determine bounds on  Planck scale interactions. 

In the last few years theoretical and phenomenological investigations have used the above effective model as an instrument to look for the Lorentz-violation from quantum field theory. The theoretical studies  are summarized in the quantization calculations \cite{Reyes1}, stability and analytical behavior  \cite{Reyes2,Scatena} and curved spacetime approach \cite{Leonardo}. On the other hand, the phenomenological studies derive constraints on Lorentz invariance violation by using recent astrophysical observations, as one can see in references \cite{maccione,gubitosi,savaliev,noordmans,Laurent,Passos2016}, for instance. Moreover, it was pointed by Brito {\textit {et al.}} \cite{brito2012} that it is also relevant to focus on low-dimensional effects of Lorentz-violating operators. Therefore, the authors successfully projected the 4D Myers-Pospelov QED in a 2D theory. The procedure of projection have led to a transition with  higher-derivative operators whose form is given as
\ben\label{proj}
&&
{\cal L}_{MP} = g\, \varepsilon^{\mu\nu\lambda\rho}n_{\mu}(n\cdot\pa)^{2}(\pa_{\nu}A_{\lambda})A_{\rho} \mapsto  \\ \nonumber
&&
{\cal L}_{2D} = g\,\varepsilon^{\alpha\,\beta}n_{\alpha}  (n\cdot\pa)^{2} \chi\,\partial_{\beta}\phi
\een
where $\alpha, \beta$ runs from $0\,\, {\rm to}\,\, 1$. One of the future perspectives pointed by \cite{brito2012} was to investigate the presence of topological defect solutions in the scalar sector of the model. Therefore, our main objective in this work is to introduce an analytical procedure to find topological defects which satisfy the equations of motion derived from the scalar sector of ${\cal L}_{\,2D}$. 

Topological defects are present in several branches of physics, such as condensed matter physics, biological models, and braneworld scenarios, for instance. 
These solutions are desired since they exhibit the property of being stable under small perturbations, 
besides they have finite energy. Some classes of models were investigated in order to search for topological 
defect solutions like those studied by Bazeia {\it et al.} \cite{bazeia2010lv}, where the authors presented 
analytical static models whose forms of the potentials and of the solutions depend on the intensity of the  
Lorentz breaking term. The previous approach was generalized by de Souza Dutra {\it et al.}\cite{dutra2011}, 
by constructing a two-field Lorentz violating model in $1+1$ dimensions,  which has analytical travelling wave and 
static defects. A relevant feature about the 2D Myers-Pospelov Lagrangian  when compared to those previous models, 
is that the first has higher-order derivative terms, which imposes several difficulties to find analytical solutions. 
In order to overcome such a problem, we introduce a procedure to solve equations of motion with higher-order derivatives. 
This procedure can be successfully applied to static and travelling wave defects. We present 
this paper in four sections which are briefly described below. 

In section \ref{model} we present the 2D Myers-Pospelov Lorentz violating 
Lagrangian for two scalar fields in $1+1$ dimensions and derive the equations of motion for such a Lagrangian for 
three different scenarios named timelike, spacelike and lightlike. Moreover in section \ref{model} we use a 
kink-like {\it Ansatz} to explore a particular version of the equations of motion, where $V=V(\phi)$. We analyze 
the features of such an {\it Ansatz} carefully for each scenario. Further, in sections \ref{asm} and \ref{atw} we 
present a procedure which  allows us to obtain analytical static and travelling wave solutions for the 
2D Myers-Pospelov model. There, we also  exemplify the procedure using a $\phi^{4}$ - $\chi^{4}$ model, and the so-called BNRT model \cite{bnrt}.  In section \ref{pertapp} we studied the behavior of the analytical defects subjected to perturbations. 
We point our conclusions, perspectives and future proposals in section \ref{remarks}.

\section{The model and some particular solutions}
\label{model}
The scalar Lagrangian modified by 2D Myers-Pospelov term can be represented as
\ben\label{v1}
{\cal L}_{2D} &=& \frac12\, \partial_{\alpha}\phi\,\partial^{\alpha}\phi + \frac12\, \partial_{\alpha}\chi\,\partial^{\alpha}\chi\ \\ \nonumber
&&
+ g\,\varepsilon^{\alpha\beta}\,n_{\alpha}(n\cdot\pa)^{2}\,\chi\,\partial_{\beta}\phi- V(\phi,\chi),
\een
where the last is written in terms of two real scalar fields, and has generalized dynamics. Here we are also considering the indexes $\alpha=\beta=0,1$, meaning a $1+1$ dimensional theory. Moreover, the constant $g$ can be called as a breaking parameter, since it
controls the influence of the Lorentz breaking term into the Lagrangian.
The equation of motion for the scalar fields $\phi$ and $\chi$ are
\ben
\Box\phi + g\epsilon^{\alpha\beta}n_{\alpha}(n\cdot\pa)^{2}\partial_{\beta}\chi + V_{\,\phi}=0,\nonumber\\
\Box\chi - g\epsilon^{\alpha\beta}n_{\alpha}(n\cdot\pa)^{2}\partial_{\beta}\phi + V_{\,\chi}=0,
\een
respectively. In the last equations, $V_{\phi}$, and $V_{\chi}$ are the partial derivatives of potential $V$ in respect to fields $\phi$, and $\chi$. Moreover, notice that the above equations depend on the choice of $n_{\alpha}$, therefore, they may change when the last one is purely timelike, is purely spacelike or is purely lightlike. The set of equations 
associated with the timelike, the spacelike, and the lightlike configurations are given by
\ben\label{tm01}
&& \slabel{tm011} \ddot{\phi} - \phi^{\prime\prime} + g \ddot{\chi}^{\prime} + V_{\,\phi}=0\,;\\&&
\slabel{tm012}\ddot{\chi} - \chi^{\prime\prime} - g \ddot{\phi}^{\prime} + V_{\,\chi}=0\,,
\een 
\ben\label{sp1}
&& \slabel{sp01}\ddot{\phi} - \phi^{\prime\prime} + g \dot{\chi}^{\prime\prime} + V_{\,\phi}=0\,;\\&&
\slabel{sp02}\ddot{\chi} - \chi^{\prime\prime} - g \dot{\phi}^{\prime\prime} + V_{\,\chi}=0\,,
\een 
\ben\label{ll1}
&& \slabel{ll01}\ddot{\phi} - \phi^{\prime\prime} + g \big(\dddot{\chi} +3\,\ddot{\chi}^{\prime} +3\, \dot{\chi}^{\prime\prime} + \chi^{\prime\prime\prime}\big)+ V_{\,\phi}=0\,;\\&&
\slabel{ll02}\ddot{\chi} - \chi^{\prime\prime} - g \big(\dddot{\phi} +3\, \ddot{\phi}^{\prime} +3\, \dot{\phi}^{\prime\prime} + \phi^{\prime\prime\prime}\big)+ V_{\,\chi}=0\,,
\een
respectively.
Let us examine in details some peculiarities about these three previous scenarios, taking a 
special case where the potential $V=V(\phi)$, which allows us to rewrite the equations of motion \eqref{tm012}, \eqref{sp02}, and \eqref{ll02} relative to the field $\chi$ in the following forms
\be \label{tm2}
\ddot{\chi}-\chi^{\,\prime\prime}-g\,\ddot{\phi}^{\,\prime}=0\,; \qquad \mbox{timelike}\,,
\ee
\be \label{sp2}
\ddot{\chi}-\chi^{\,\prime\prime}-g\,\dot{\phi}^{\,\prime\prime}=0\,; \qquad \mbox{spacelike}\,,
\ee
\be \label{ll2}
\ddot{\chi}-\chi^{\,\prime\prime}-g\,\left(\dddot{\phi}+3\,\ddot{\phi}^{\,\prime}+3\,\dot{\phi}^{\,\prime\prime}+\phi^{\,\prime\prime\prime}\right)=0\,; \qquad \mbox{lightlike}\,.
\ee
Then, if we consider 
\be \label{ans1}
\phi(x,t)=\tanh\,(k\,x+\omega\,t)\,,
\ee
\noindent as an {\it Ansatz} for the field $\phi$, we can clearly see that the relations
\ben \label{ans2}
&&
\dot{\phi}=\omega\,(1-\phi^2)\,; \qquad  \phi^{\,\prime}=k\,(1-\phi^{\,2})\,; \qquad  \dot{\phi}^{\,\prime}=-2\,\omega\,k\,\,(1-\phi^{\,2})\,\phi\\ \nonumber
&&
\ddot{\phi}=-2\,\omega^{\,2}\,(1-\phi^{\,2})\,\phi\,; \qquad  \phi^{\,\prime\,\prime}=-2\,k^{\,2}\,(1-\phi^{\,2})
\,\phi\,;  \\ \nonumber
&&
\ddot{\phi}^{\,\prime}=6\,\omega^2\,k\,\phi^{\,2}\,(1-\phi^{\,2})-2\,\omega^2\,\phi^{\,\prime}\,; \qquad \dot{\phi}^
{\,\prime\prime}=6\,\omega\,k^{\,2}\phi^{\,2}(1-\phi^{\,2})-2\,\omega\,k\,\phi^{\,\prime}\,; \\ \nonumber
&&
\dddot{\phi}=6\,\omega^{\,3}\,\phi^{\,2}\,(1-\phi^{\,2})-2\,\omega^2\,\dot{\phi}\,;\qquad  \phi^{\,\prime\prime\prime}=6\,k^{\,3}\,\phi^{\,2}\,(1-\phi^{\,2})-2\,k^2\,\phi^{\,\prime}\,,
\een

\noindent are satisfied for such a solution. Now, let us take 
\be \label{def1}
\chi=\phi^{\,2}/2\,,
\ee
as a connection between both fields, which can be applied together with the equations 
presented in \eqref{ans2} to analyze each different configuration. Starting with the timelike scenario, 
the previous ingredients lead us to rewrite \eqref{tm011} as
\be \label{b01}
-\phi^{\,\prime\,\prime}-8\,g\,\omega^{\,2}\,\phi\,\phi^{\,\prime}=0\,,
\ee
for
\be
V_{\,\phi}(\phi)=2\,\omega^{\,2}(1-\phi^{\,2})\,\phi-12\,g\,\omega^{\,2}\,k\,(1-\phi^{\,2})\,\phi^{\,3}\,,
\ee
where via dimensional analyses, we can note that $g$ has dimension of $k^{\,-1}$, and $\omega$ has dimension of $k$.
Therefore, taking our initial {\it Ansatz} into Eq. \eqref{b01}, yields us to determine the dispersion relation $k=4\,g\,\omega^2$. Moreover, Eq. \eqref{tm2} imposes that $\omega_{\pm}=\pm\,\left(2\,\sqrt{2}\,g\right)^{\,-1}$, which means that $k=(2\,g)^{\,-1}$. So, the final form of our potential is simply given by
\be
V(\phi)=\frac{\phi ^2}{8 g^2} \, \left(1-\phi ^2\right)^2\,.
\ee
In the spacelike regime, Eq. \eqref{sp01} is represented as
\be \label{b02}
-\phi^{\,\prime\,\prime}-8\,g\,\omega\,k\,\phi\,\phi^{\,\prime}=0\,,
\ee
if
\be
V_\phi(\phi)=2\,\omega^{\,2}(1-\phi^{\,2})\,\phi-12\,g\,\omega\,k^{\,2}\,(1-\phi^{\,2})\,\phi^{\,3}\,,
\ee
such results impose the constraint $\omega=(4\,g)^{\,-1}$ for the solution \eqref{ans1}. Consequently, Eq. \eqref{sp2} implies that $k_{\pm}=\,\pm(2\,\sqrt{2}\,g)^{\,-1}$, and the final form for our potential in this case is
\be
V(\phi)=\frac{\phi ^2}{16 g^2}\, \left(1-\phi ^2\right)^2\,.
\ee

After several manipulations based on \eqref{ans2}, and \eqref{def1}, the lightlike equation of motion for field $\phi$ can be reduced to
\be \label{b03}
-\phi^{\,\prime\,\prime}-2\,g\,\left(4\,k^{2}+3\,\omega^{2}+3\,\omega\,k\right)\,\phi\,\phi^{\,\prime}=0\,,
\ee
for
\ben \label{b04}
V_{\phi}&=&2\, \phi\,  \left(\phi^2-1\right) \big[6\, g\, k^3\, \phi^2+9 \,g\, k^2\, \omega\,  \left(2\, \phi^2-1\right) \\ \nonumber
&&
+\omega^2 \left(9\, g\, k \left(2\, \phi^2-1\right)-1\right)+2\,g\, \omega^3 \left(3 \,\phi^2-2\right)\big]\,.
\een
So, by inspecting Eq. \eqref{b03} subject to \eqref{ans1} we find the dispersion relation 
\be \label{b05}
k_{\pm}=\frac{1-3\,g\,\omega \pm \sqrt{1-39\,g^2\,\omega^2-6\,g\,\omega}}{8\,g}\,,
\ee
moreover, Eq. \eqref{ll2} results in
\be \label{b06}
k=-\omega\,; \qquad k_{\pm}=\frac{1-4 g \omega \pm \sqrt{1-16 g \omega }}{4 g}\,.
\ee
Thus, Eqs. \eqref{b05}, and \eqref{b06} lead us to the following constraints for $\omega$: $\omega_{1+}=-(4\,g)^{\,-1}$, $\omega_{1-}=0$, $\omega_{2+}=\,(16\,g)^{\,-1}$, and $\omega_{2-}=0$. Substituting such constraints back in \eqref{b05}, we find  $k_{1+}=\,(4\,g)^{\,-1}$, $k_{1-}=0$, $k_{2+}=3\,(16\,g)^{\,-1}$, and $k_{2-}=0$. The last results imply in $V_{1\,\phi}=0$, and 
\be
V_{2\,\phi}=\frac{3\,\phi^5-4\,\phi^3+\phi}{16\,g^2}\,,
\ee
respectively. Thus, the potential $V_{2}$ can be written as
\be
V_2(\phi)=\frac{\phi^2}{32\,g^2}\, \left(1-\phi^2\right)^2\,.
\ee
An interesting feature in respect to this approach is that Eqs.
\eqref{b01}, \eqref{b02}, and \eqref{b03} can be viewed as the famous Burgers equation \cite{burgers1,burgers2}
for waves with low velocities.

\begin{figure}[h!]
\centering
\hspace{0.2cm}
\includegraphics[width=0.4\columnwidth]{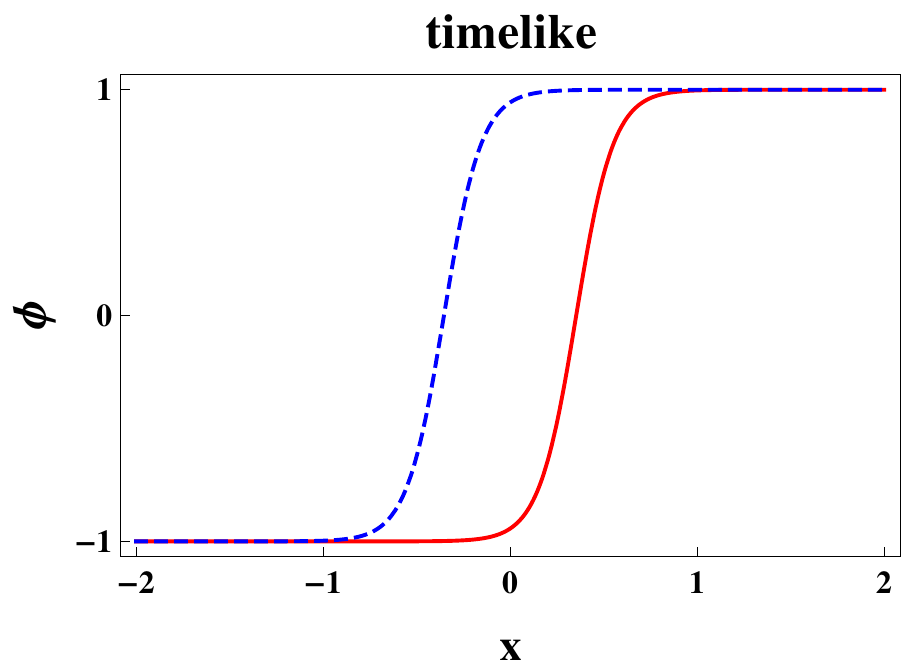} \hspace{0.2 cm} \includegraphics[width=0.4\columnwidth]{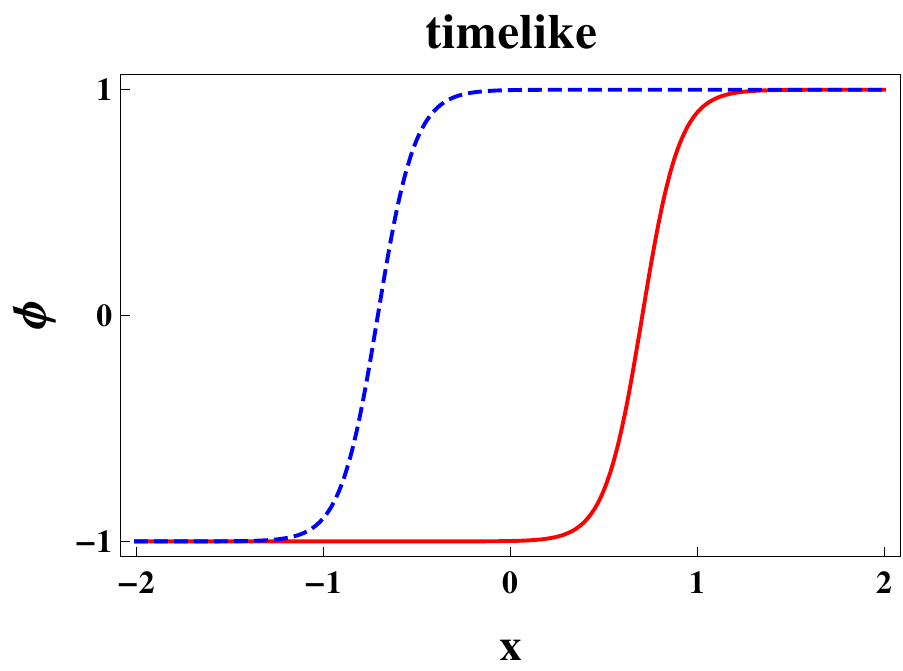}

\hspace{0.2 cm} \includegraphics[width=0.4\columnwidth]{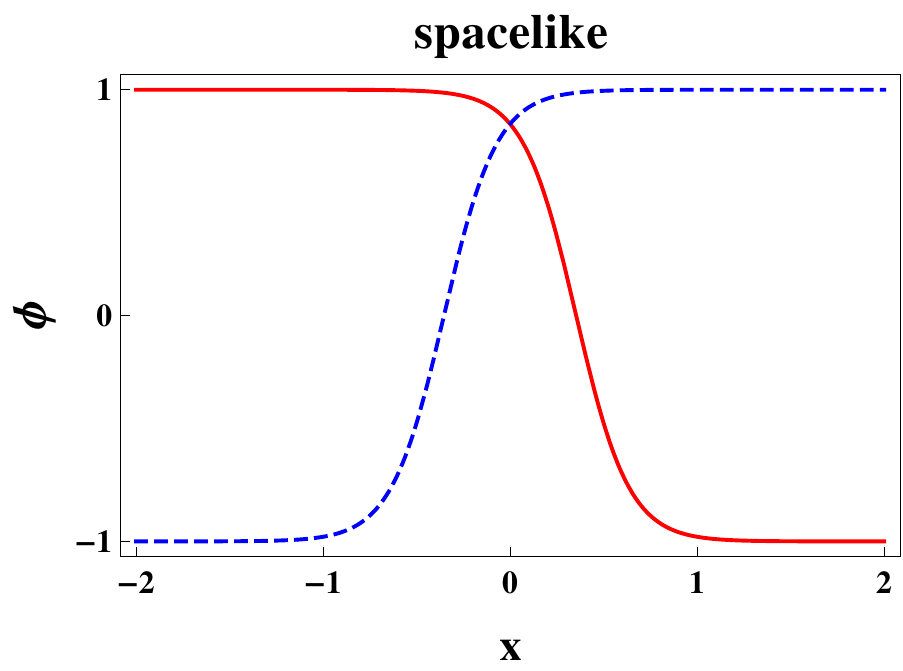} \hspace{0.2 cm} \includegraphics[width=0.4\columnwidth]{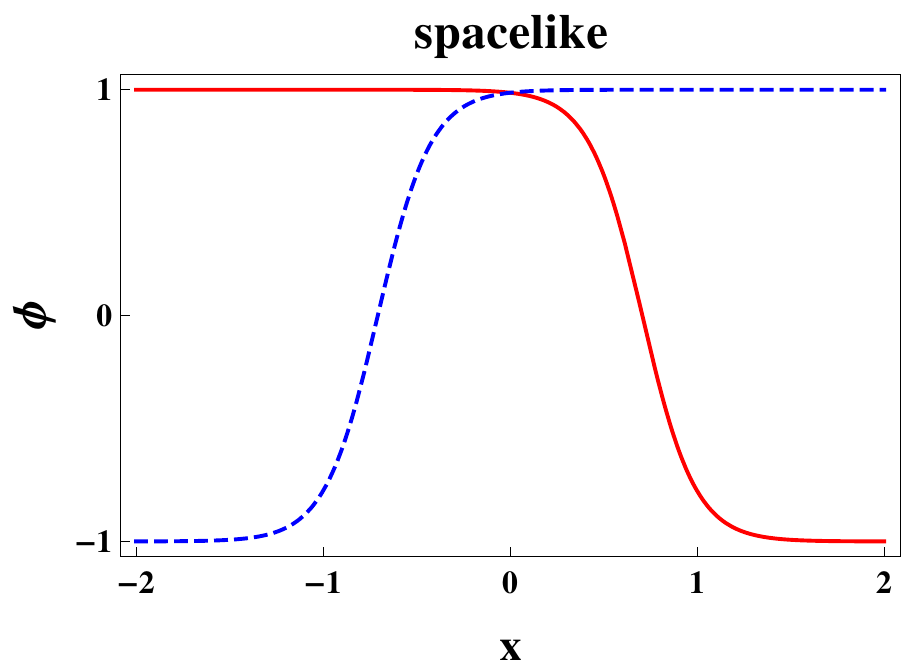}

\vspace{0.2cm}
\hspace{0.2 cm} \includegraphics[width=0.4\columnwidth]{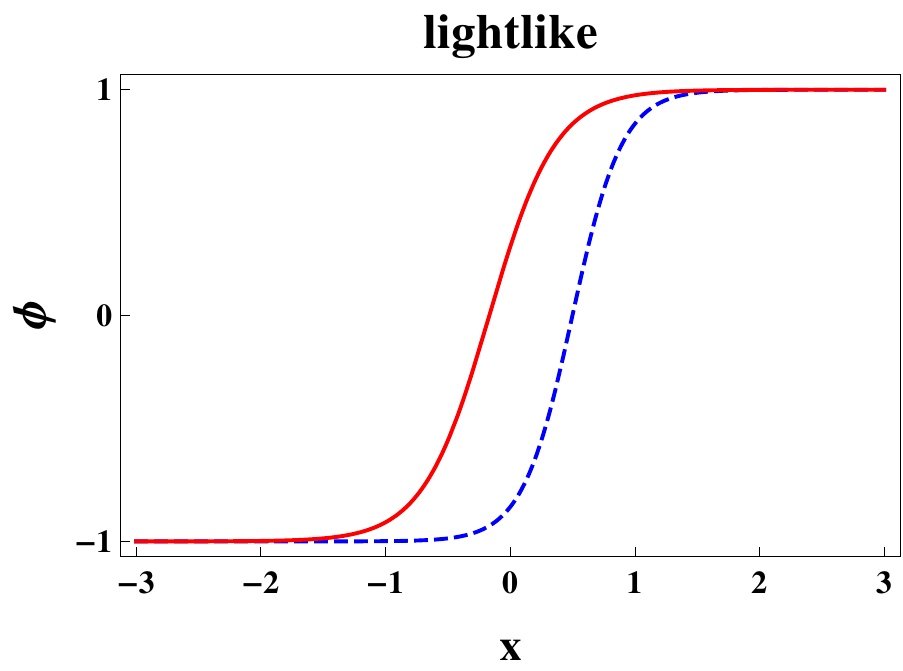} \hspace{0.2 cm} \includegraphics[width=0.4\columnwidth]{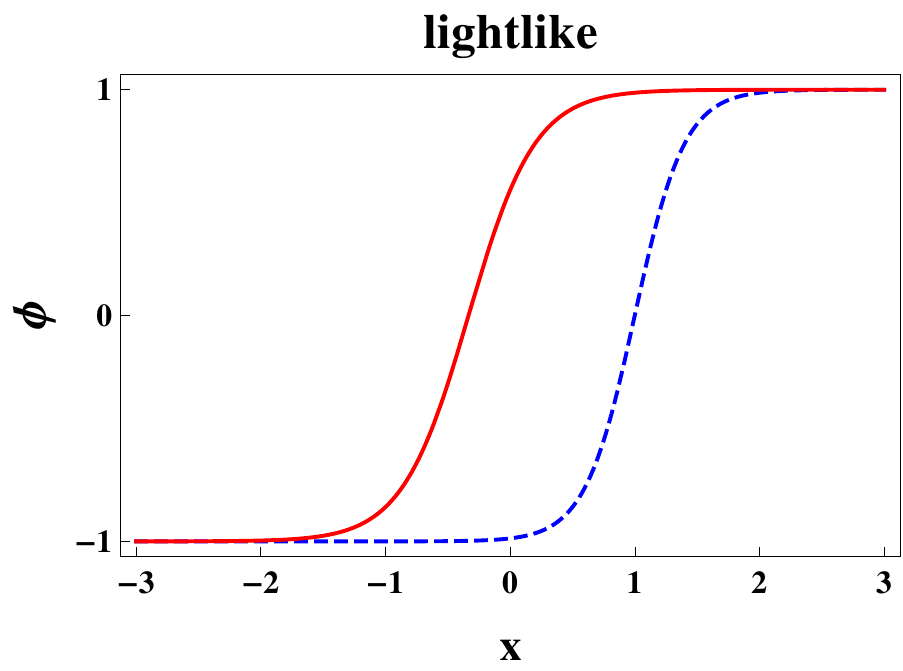}
\caption{Field $\phi$ Eq. \eqref{ans1} for timelike, spacelike and lightlike scenarios, with $g=0.1$. The time evolution of the solutions unveil the chiral behavior of the spacelike and of the lightlike solutions.}
\label{FIG0}
\end{figure}

Some solutions which satisfy Burgers equation (as well as, its
generalized versions such those introduced in \cite{bazeia98}), can present 
the nice property of being chiral, which means that the waves travel with well defined signs for their 
velocities \cite{bazeia98}. This chiral propagation is observed in the spacelike, and in the lightlike solutions as we can see in Fig \ref{FIG0}. The graphs were plotted with $t=0.5$ (left-hand side of each scenario) and $t=1$ (right-hand side of each scenario), there the blue (dashed) curves show $\phi$ for $k$ and $\omega_+$, for $k_+$ and $\omega$, and for $k_{1}$ and $\omega_1$ for  timelike, spacelike and lightlike scenarios, respectively. Moreover, the red (solid) curves represent   $\phi$ for $k$ and $\omega_-$, for $k_-$ and $\omega$, and for $k_{2}$ and $\omega_2$ for  timelike, spacelike and lightlike cases, respectively. 
It is also interesting to note that the dispersion relations founded show that $k$ and $\omega$ depend on the breaking parameter $g$, 
which means that the fields $\phi$ and $\chi$ are sensible to the Lorentz violation. 

\section{General analytical static models}
\label{asm}

Despite the previous solutions that we found, it is relevant to search for general sets of fields which can satisfy the equations of motion of the three presented configurations in a more natural way. In order to do so, let us firstly work with one-dimensional static defects, which means that $\phi=\phi(x)$ and $\chi=\chi(x)$. Such a choice results in the following forms for Eqs. \eqref{tm01} and \eqref{sp1}
\be \label{am_1}
-\phi^{\,\prime\prime}+V_{\,\phi}=0\,; \qquad -\chi^{\,\prime\prime}+V_{\,\chi}=0\,,
\ee
and these are the equations of motion related with a standard two-field Lagrangian, which can present stable  solutions. However, the set of equations of motion mentioned in \eqref{ll1} is rewritten as
\ben \label{am_2}
&&
- \phi^{\prime\prime} + g \,\chi^{\prime\prime\prime}+ V_{\,\phi}=0\nonumber\\
&&
- \chi^{\prime\prime} - g \,\phi^{\prime\prime\prime}+ V_{\,\chi}=0\,,
\een 
consequently, this set stills preserve general kinetic terms. 

The previous set of differential equations can be integrated if we multiply the first one by $\phi^{\,\prime}$ and the second one by $\chi^{\,\prime}$, then, by adding both expressions we find 
\be \label{am_3}
-\phi^{\,\prime\prime}\,\phi^{\,\prime}-\chi^{\,\prime\prime}\,\chi^{\,\prime}+g\,\chi^{\,\prime\prime\prime}\,\phi^{\,\prime}-g\,\phi^{\,\prime\prime\prime}\,\chi^{\,\prime}+V_{\,\phi}\,\phi^{\,\prime}+V_{\,\chi}\,\chi^{\,\prime}=0\,,
\ee
so, we directly determine that potential $V$ has the form
\be \label{am_4}
V(\phi,\chi)=\frac{\phi^{\,\prime\,2}}{2}+\frac{\chi^{\,\prime\,2}}{2}-g\,\left(\chi^{\,\prime\prime}\,\phi^{\,\prime}-\phi^{\,\prime\prime}\,\chi^{\,\prime}\right)\,,
\ee
 which was obtained by integration of \eqref{am_3} in respect to $x$.
The previous reduction process lead us to a potential with depends on first and second order derivatives of fields $\phi$, and $\chi$. However, such a process does not give us any clue about a proper {\it Ansatz} for a first-order formalism. As an inspiration to establish proper forms for the first-order differential equations related with our system, let us review the Lorentz breaking model studied by Barreto {\it et al.} in \cite{bbm}. There, the authors solved the following equations of motion
\be \label{am_5}
\phi^{\,\prime\prime}+g\,\chi^{\,\prime}=V_{\,\phi}\,; \qquad \chi^{\,\prime\prime}-g\,\phi^{\,\prime}=V_{\,\chi}\,,
\ee
so by multiplying the left hand side equation by $\phi^{\,\prime}$, and the right hand side one by $\chi^{\,\prime}$, and adding both expressions, we find
\be
\phi^{\,\prime\prime}\,\phi^{\,\prime}+\chi^{\,\prime\prime}\,\chi^{\,\prime}=V_{\,\phi}\,\phi^{\,\prime}+V_{\,\chi}\,\chi^{\,\prime}\,.
\ee
Therefore, the integration of the last equation in respect to $x$ yields to the potential
\be  \label{am_6}
V=\frac{\phi^{\,\prime\,2}}{2}+\frac{\chi^{\,\prime\,2}}{2}\,.
\ee

In their investigation, Barreto {\it et al.} worked  with the first-order differential equations
\be  \label{am_7}
\phi^{\,\prime}=W_{\,\phi}(\phi,\chi)+s_1\,\chi\,; \qquad \chi^{\,\prime}=W_{\,\chi}(\phi,\chi)+s_2\,\phi\,,
\ee
as a correct {\it Ansatz} to solve \eqref{am_5}. Here $s_1$, and $s_2$ are real constants, besides $W_{\phi}$, and $W_{\chi}$ are the derivatives of a superpotential $W(\phi,\chi)$ in respect to $\phi$, and $\chi$. Therefore, substituting \eqref{am_7} into \eqref{am_5}, and \eqref{am_6}, one can compute that their {\it Ansatz} satisfy the equations of motion if $g=s_2-s_1$. Let us analyze carefully what is the specific reason behind the first-order formalism proposed in \cite{bbm}. For such an approach, we work with the generic forms
\be \label{am_8}
\phi^{\,\prime}=F(\phi,\chi)\,; \qquad \chi^{\,\prime}=G(\phi,\chi)\,,
\ee
so, by putting these first-order equations into  \eqref{am_5}, and \eqref{am_6} we establish the constraint
\be \label{am_9}
g+F_{\,\chi}-G_{\,\phi}=0\,,
\ee
thus, if we take $F=W_{\,\phi}+s_1\,\chi$, and $G=W_{\,\chi}+s_2\,\phi$, \eqref{am_9} lead us to $g=s_2-s_1$, as required. 

Consequently, the major constraint for the first-order formalism related with the Lorentz breaking model introduced in \cite{bbm}, is the one described in Eq. \eqref{am_9}. Analogously, the application of {\it Ansatz} \eqref{am_8} in \eqref{am_2}, and \eqref{am_4} results in the constraint
\ben \label{am_10}
&&
G_{\,\phi}+F_{\,\chi}\,\left(3\,g\,G_{\,\phi}-1\right)+g\,\left[G_{\,\chi}^{\,2}-G_{\,\chi}\,F_{\,\phi}+F_{\,\phi}^{\,2}\right.\\ \nonumber
&&
\left.+G\,\left(G_{\,\chi\chi}+F_{\,\phi\chi}\right)+F\,\left(G_{\,\phi\chi}+F_{\,\phi\,\phi}\right)\right]=0\,,
\een
note that this difference between \eqref{am_10}, and \eqref{am_9} is a direct consequence of terms $g\,\phi^{\,\prime\prime\prime}$, and $g\,\chi^{\,\prime\prime\prime}$ presented in our equations of motion. 
One possibility to solve the previous equation for models which have defects solutions, is based on the following forms for $F$ and $G$
\ben \label{am_11}
&&
F=W_{\,\phi}(\phi,\chi)+Z(\phi)\,\left[\phi-f(\chi)\right]\,; \qquad \phi-f(\chi)=r\,; \\ \nonumber
&&
G=W_{\,\chi}(\phi,\chi)+Z(\chi)\,\left[\phi-f(\chi)\right]\,; \qquad  Z(\phi)=Z(\chi)\,,
\een
here, $W_\phi$, and $W_\chi$ are the derivatives of a superpotential $W$ in respect to fields $\phi$, and $\chi$. Besides, $W$ is related with an analytical standard two-field theory. 

{We also have $Z$, which is an arbitrary function whose explicit form is determined by solving \eqref{am_10}. Moreover, we use the mapping (or the orbit equation \cite{orb1}), $\phi-f(\chi)=r$ to write $Z(\phi)$ as $Z(\chi)$. So, for a given connection between $\phi$ and $\chi$, the terms multiplying $Z$ in \eqref{am_11} evaluate to $r$. Furthermore, once we have a form for function $Z$, the solutions of the first-order differential equations are going to automatically satisfy the equations of motion presented in \eqref{am_2}. }

\subsection{Example - $\phi^{\,4}$ - $\chi^{\,4}$ with $r=0$}

\begin{figure}[ht!]
\centering

\includegraphics[width=0.5\columnwidth]{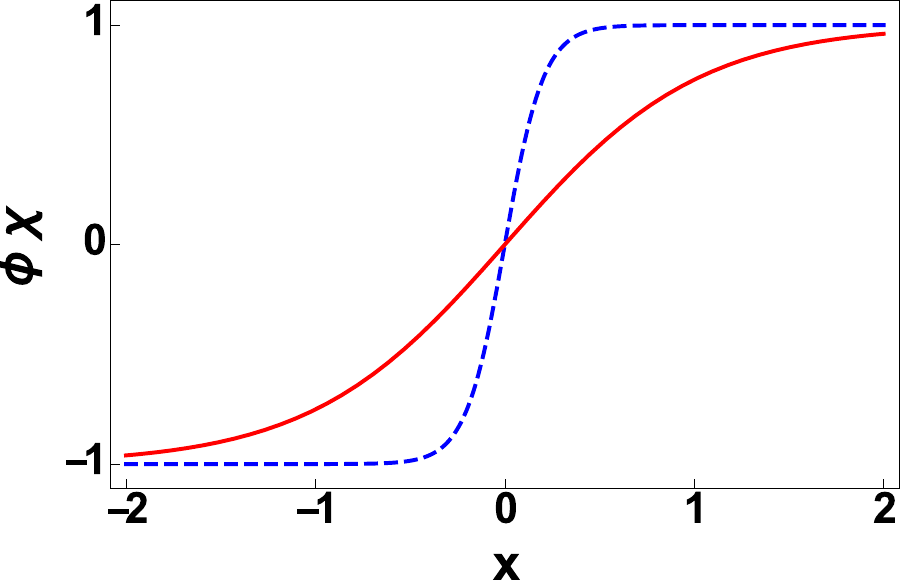}
\vspace{0.3cm}
\caption{Fields $\phi$, and $\chi$ for the $\phi^{\,4}$ - $\chi^{\,4}$ model. The kinks were generated with $k=(\alpha\,g)^{\,-1}$, $\alpha = 204.96$, $g=0.005$ (solid red curve), and $g=0.001$ (dashed blue curve). It is possible to realize that smaller values of $g$ result in more intense transitions between the vacua.}
\label{FIG1}
\end{figure}

Let us illustrate the methodology of the last section with an example involving a $\phi^{\,4}$ plus a $\chi^{\,4}$ model, which is derived from the superpotential 
\be \label{ex_1}
W=k\,\left(\phi-\frac{\phi^{\,3}}{3}\right)+k\,\left(\chi-\frac{\chi^{\,3}}{3}\right)\,.
\ee
In this approach we are going to set $r=0$ for a matter of simplicity. From Eq. \eqref{am_11} we can observe that the previous superpotential results in the respective forms for functions $F$, and $G$ 
\be \label{ex_4}
F=k\,\left(1-\phi^{\,2}\right)+Z(\phi)\,(\phi-f(\chi))\,; \qquad G=k\,\left(1-\chi^{\,2}\right)+Z(\chi)\,(\phi-f(\chi))\,.
\ee
Both functions need to be substituted in Eq. \eqref{am_10} in order to compute $Z$. Such a procedure yields to the constraint
\ben \label{ex_4_1}
&& \nonumber
k \left(1-\chi ^2\right) \bigg\{f_\chi\,Z (1-3 g Z)+g\, \bigg[k \left(1-f^2\right) \left(\left(\frac{2}{f_\chi}+1\right) Z_{\chi}-2 \,k\right) \\ 
&&
+\left(f_\chi\,Z+2\, k\, \chi \right)^2-(2\,k f-Z)\, \left(f_\chi\,Z+2\, k\, \chi \right)\\ \nonumber
&&
+k \left(\chi ^2-1\right) \left( f_{\chi\,\chi}\,Z+\left(2 f_\chi+1\right)\, Z_{\chi}+2 k\right)+(Z-2 \,k\, f)^2\bigg]+Z\bigg\}=0\,,
\een
where $Z=Z(\chi)$, and we worked with
\be \label{ex_4_2}
Z(\phi)=Z(\chi)\,; \qquad Z_{\,\phi}=\frac{Z_{\chi}}{f_{\chi}}\,; \qquad Z_{\phi\,\phi}=\frac{Z_{\chi\,\chi}}{f_{\chi}^{\,2}}-\frac{f_{\,\chi\chi}}{f_{\chi}^{\,3}}\,Z_\chi\,.
\ee
Then, from the entire set of orbits connecting fields $\phi$ and $\chi$ for the superpotential $W$, the one which mostly simplifies the constraint \eqref{ex_4_1} is 
\be \label{ex_3}
\phi-\chi=0\,; \qquad f(\chi)=\chi\,; \qquad f_{\chi}=1\,; \qquad f_{\chi\,\chi}=0\,,
\ee
corresponding to a straight line orbit in the $\phi-\chi$ plane. Such an assumption results in the following form for $Z(\chi)$
\be \label{ex_5}
Z(\chi)=2\,g\, k^2\, \left(1-2 \,\chi ^2\right)\,.
\ee
Thus, the first-order approach for this Lorentz breaking model is given by
\ben \label{ex_6}
&&
\phi^{\,\prime}=k\,\left(1-\phi^{\,2}\right)+2\,g\, k^2\, \left(1-2 \,\phi ^2\right)\,(\phi-\chi)\,; \\ \nonumber
&&
\chi^{\,\prime}=k\,\left(1-\chi^{\,2}\right)+2\,g\, k^2\, \left(1-2 \,\chi ^2\right)\,(\phi-\chi)\,,
\een
whose analytical solutions are
\be \label{ex_2}
\phi(x)=\tanh(k\,x)\,; \qquad \chi(x)=\tanh(k\,x)\,,
\ee
unveiling defects which exhibit a kink-like profile, as one can see in Fig. \ref{FIG1}.

These first-order differential equations can be used together with \eqref{am_4} to build the effective two scalar field potential $V(\phi,\chi)$, yielding to

\ben \label{ex_7} \nonumber 
V(\phi,\chi)&=&\frac{k^2}{2} \,\bigg[64\,g^4\, k^4\, \left(1-2\, \chi ^2\right)\, \left(2\, \phi ^2-1\right)\, (\phi -\chi )^4-16\, g^3\, k^3 \,(\phi -\chi )^3\, \\ \nonumber
&&
\times\,\left(5+3\, \left(4\, \chi ^2-3\right)\, \phi ^2-2\, \chi \, \phi-9\, \chi ^2 \right) +\,4\, g^2\, k^2\, (\phi -\chi )^2 \\ \nonumber
&&
\times\, \left(4\, \chi ^4+7\, \chi ^2+4\, \phi ^4+\left(7-12\, \chi ^2\right)\, \phi ^2+2\, \chi\,  \phi -6\right) \\ \nonumber
&&
+\,4\, g\, k\, (\phi -\chi )\, \left(2\, \chi ^4-2\, \chi ^2+2\, \phi ^4-\left(\chi ^2+2\right)\, \phi ^2+1\right) \\ 
&&
+\chi ^4-2\, \chi ^2+\phi ^4-2\, \phi ^2+2\bigg]\,.
\een
With the last equation in hands, we are able to derive $V_\phi$, and $V_\chi$, which in turn can be used to prove that the analytical solutions from \eqref{ex_2} satisfy the equations of motion \eqref{am_2}. Therefore, despite the complexity of the previous potential, our methodology can be easily applied to find an analytical model in this Lorentz-breaking scenario.

The total energy of the defects is computed via the integration of the Hamiltonian density over the entire space, which for such a system is derived from the zero component of the energy-momentum tensor, whose form for this specific dynamics is \cite{rego-monteiro13}
\ben \label{ex_7_1}
&&
\hspace{-0.5cm}H=T^{\,0\,0} \\ \nonumber
&&
\hspace{-0.5cm}=\sum_{i=1}^{2}\,\bigg\{\bigg[\frac{\partial{\cal L}}{\partial\,(\partial_0\,\phi_{\,i})} -\partial_{\,\alpha}\,\frac{\partial\,{\cal L}}{\partial\,(\partial_{\,0}\partial_{\,\alpha}\,\phi_{\,i})}\bigg]\,\partial^{0}\,\phi_{i}
+\bigg[\frac{\partial\,{\cal L}}{\partial\,(\partial_{\,0}\,\partial_{\,\alpha}\,\phi_{\,i})}\bigg]\,\partial_{\,\alpha}\,\partial^{0}\,\phi_{\,i}\bigg\} -{\cal L}\,;  \\ \nonumber
&&
\hspace{-0.5cm}\phi_{1}=\phi\,; \qquad \phi_{2}=\chi\,.
\een

 Besides, the Hilbert stress-energy tensor represents an alternative route to determine the Hamiltonian density. Such a tensor is given by
\be \label{ex_7_2}
T^{\,\mu\,\nu}=\frac{2}{\sqrt{-g}}\,\frac{\delta\,(\sqrt{-g}\,{\cal L})}{\delta\,g_{\,\mu\,\nu}}=2\,\frac{\delta\,{\cal L}}{\delta\,g_{\,\mu\,\nu}}+\frac{2}{\sqrt{-g}}\,\frac{\delta\,\sqrt{-g}}{\delta\,g_{\,\mu\,\nu}}\,{\cal L}\,,
\ee
where
\be \label{ex_7_3}
\delta\,{\cal L}=\frac{\partial\,{\cal L}}{\partial\,g_{\,\mu\,\nu}}\,\delta\,g_{\,\mu\,\nu}\,; \qquad \delta(\sqrt{-g})=-\frac{1}{2}\,\sqrt{-g}\,g^{\,\mu\,\nu}\,\delta\,g_{\,\mu\,\nu}\,.
\ee
The previous relations allow us to rewrite $T^{\,\mu\,\nu}$ as
\be \label{ex_7_4}
T^{\,\mu\,\nu}=2\,\frac{\partial\,{\cal L}}{\partial\,g_{\,\mu\,\nu}}-g^{\,\mu\,\nu}\,{\cal L}\,.
\ee
Consequently if the defects are statical, $T^{\,0\,0}$ is simply
\be \label{ex_7_5}
H=T^{\,0\,0}=-{\cal L}\,,
\ee
for Minkowski space-time. This form of $H$ is consistent with the statical version of Eq. $(\ref{ex_7_1})$.

As we are dealing with static defects, the previous equation yields to $H=-{\cal L}$, where ${\cal L}$ is explicitly described by
\be \label{ex_7_6}
{\cal L}=-\frac{\phi^{\,\prime\,2}}{2}-\frac{\chi^{\,\prime\,2}}{2}+g\,\chi^{\,\prime\,\prime}\,\phi^{\,\prime}-V(\phi,\chi)\,,
\ee
with $V$ presented in Eq. \eqref{ex_7}. Consequently, our Hamiltonian density for these static defects is
\be \label{ex_7_12}
H=2\, k^2\, \text{sech}^4(k x) \left[g\, k\, \tanh (k x)+1\right]\,,
\ee
resulting in the total energy
\be \label{ex_7_13}
E=2\, k^2\,\int_{-\infty}^{+\infty}\,d\,x\, \text{sech}^4(k x) \left[g\, k\, \tanh (k x)+1\right]=\frac{8}{3}\,k\,.
\ee
We can also point that the Hamiltonian density $(\ref{ex_7_12})$ as well the potential $V$ $(\ref{ex_7})$, are not symmetric under the exchange of $x$ by $-x$ (parity transformation), and of $\phi$ by $\chi$, respectively. These symmetries are broken as long as $g$ is different of zero, representing a direct consequence of the Lorentz-breaking process. In such cases, the Lorentz-breaking parameter imposes different energy densities on either sides of the domain-walls. 

The introduced procedure enable us to find several analytical models for a Lorentz breaking theory with $k$, and $g$ independent, however not all of these models are going to be physically interesting. Therefore, it is crucial to establish a constraint between $k$, and $g$ to determine proper forms for $V$. 

A dependence between $k$ and $g$ can be determined through an analysis of the extreme points of potential $V$. It is possible to show that this potential has a local maximum at $(0,0)$, besides two minima at $(1,1)$, and $(-1,-1)$. By computing  $V_{\,\phi\,\phi}$, and the determinant of the Hessian matrix $D=V_{\phi\phi}\,V_{\chi\chi}-V_{\phi\chi}^{\,2}$, at vacua $\phi_v=\chi_v=\pm\,1$, we observe that they are going to be truly minima of $V$ if 
\be \label{ex_8}
2\,g\,k\, \left(1-3\, g\, k\right)<1\,,
\ee
leading us to a first constraint between $k$ and the Lorentz breaking parameter $g$. 

\begin{figure}[h!]

\includegraphics[width=0.45\columnwidth]{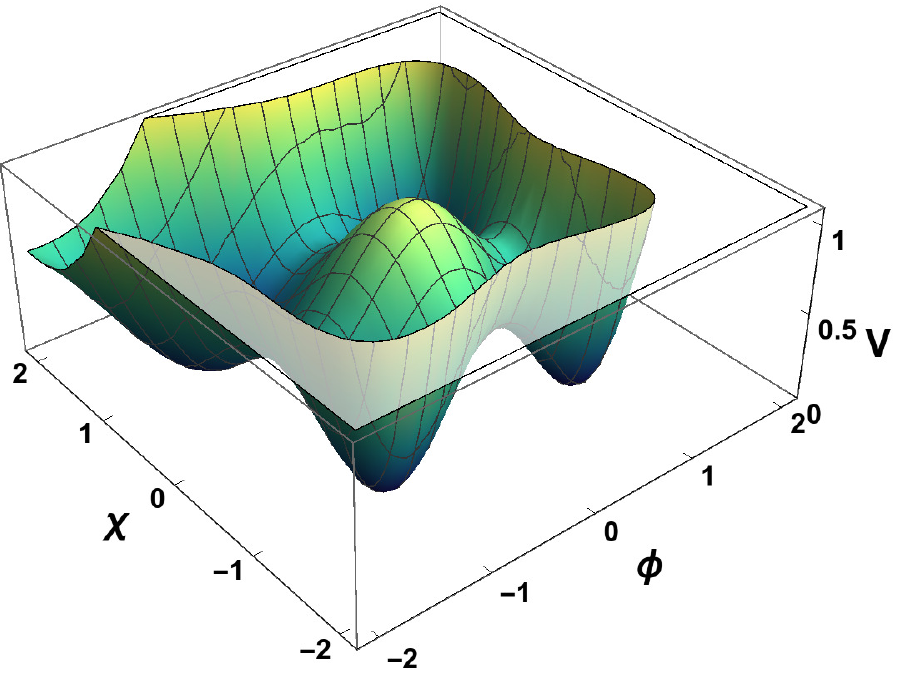} \hspace{0.2 cm} \includegraphics[width=0.45\columnwidth]{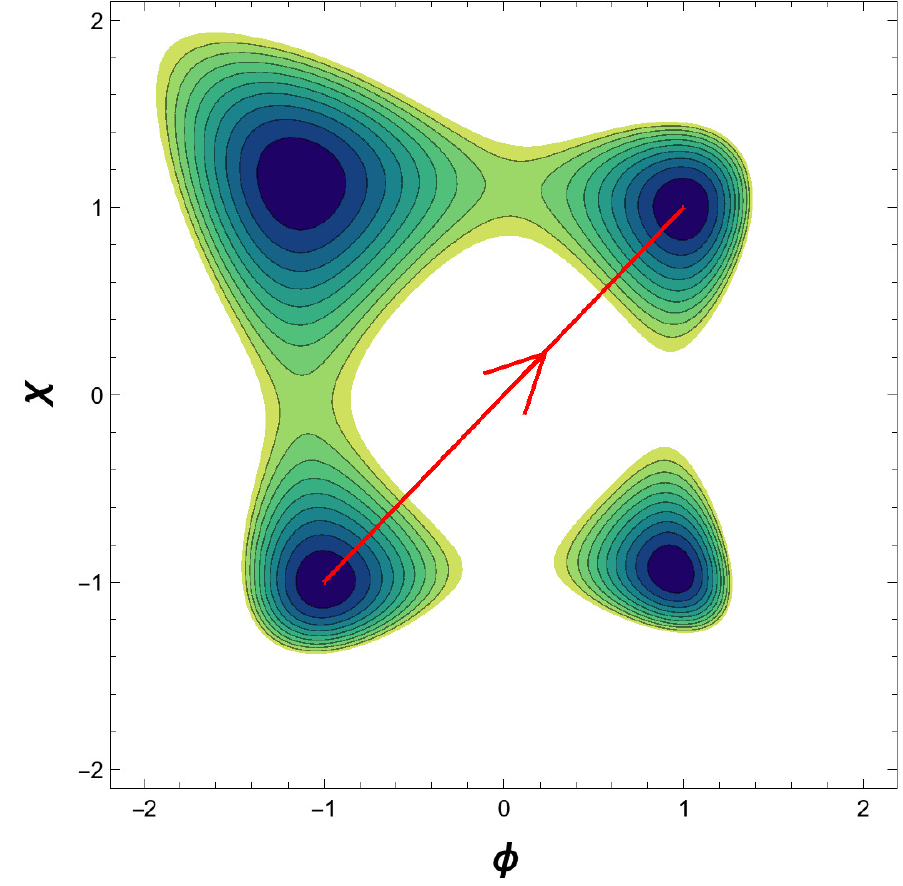}
\vspace{0.3cm}
\caption{The left panel shows potential $V$ for $k=(\alpha\,g)^{\,-1}$, $\alpha=24.61$, and $k=1$. The right graphic exhibits the contour of $V$, where the red line represents the analytical straight line orbit of our model connecting the minimum $(-1,-1)$ with $(1,1)$. }
\label{FIG2}
\end{figure}

\begin{figure}[ht!]

\includegraphics[width=0.45\columnwidth]{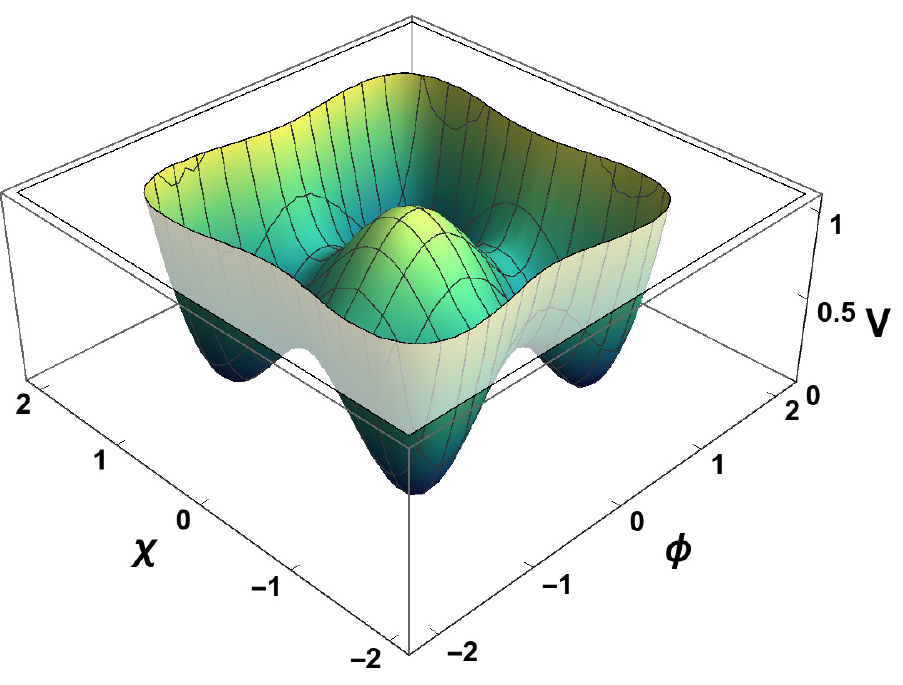} \hspace{0.2 cm} \includegraphics[width=0.45\columnwidth]{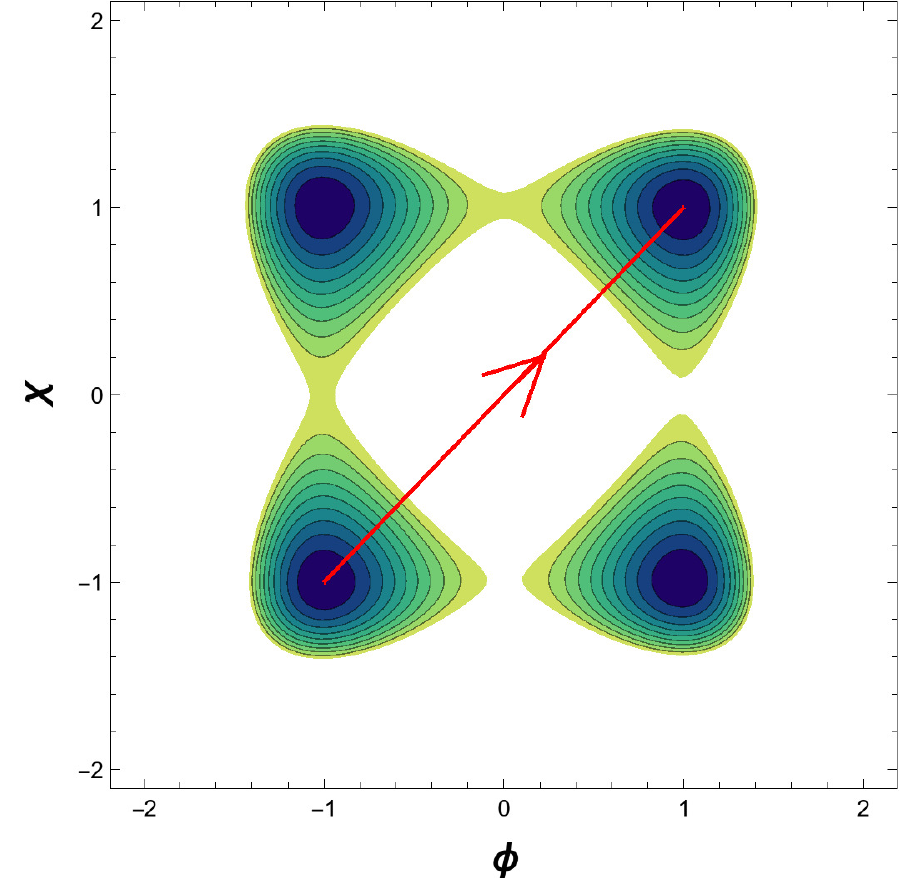}
\vspace{0.3cm}
\caption{The left panel shows potential $V$ for $k=(\alpha\,g)^{\,-1}$, $\alpha=204.96$, and $k=1$. The right graphic exhibits the contour of $V$, where the red line represents the analytical straight line orbit of our model connecting the minimum $(-1,-1)$ with $(1,1)$. }
\label{FIG3}
\end{figure}

Furthermore, if we take $g=0$, the potential $V$ reduces to its standard form given by
\be \label{ex_9}
V_{\,s}(\phi,\chi)=\frac{k^2}{2}\,  \left(\chi ^4-2\, \chi ^2+\phi ^4-2\, \phi ^2+2\right)\,,
\ee
which has four saddle points at $(\pm\,1,0)$, and $(0,\pm\,1)$, corresponding to \\ $V_{\,s}(\pm\,1,0)=V_{\,s}(0,\pm\,1)=k^2/2$. By working with $\phi=\pm\,1$ and $\chi=0$ in Eq. \eqref{ex_7}, we have 
\ben \label{ex_10}
&& \nonumber
\hspace{-0.5 cm}V(\pm\,1,0)=\frac{k^2}{2}\pm\,2\, g\, k^3+10\, g^2\, k^4 
-32\, g^3\, k^5+32\, g^4\, k^6\equiv V_{\,s}(\pm\,1,0)+c\,k^2\,; \\ 
&&
\hspace{-0.5 cm}c=\pm\,2\, g\, k+10\, g^2\, k^2-32\, g^3\, k^3+32\, g^4\, k^4\,,
\een
where $c$ should represent a small correction to the standard potential at $(\pm1,0)$ caused by the Lorentz breaking parameter, or in other words, $c$ measures how different our potential is from $V_s$ at these points. Besides, \eqref{ex_10} tells us that in the limit $g\rightarrow 0$, and $c \rightarrow 0$, we have
\be 
V(\pm\,1,0)=V_{\,s}(\pm\,1,0)\,.
\ee
Now, noting that $g$ has dimension $k^{\,-1}$, let us define that $k \equiv (\alpha\,g)^{\,-1}$, where $\alpha$ is a real nondimensional constant, therefore, Eq. \eqref{ex_10} is rewritten as
\be \label{ex_11}
\frac{(\alpha \pm 2)^2\, \left(\alpha ^2+16\right)}{2\, \alpha ^4} = \left(\frac{1}{2}+c\right)\,.
\ee
In order to make sure that $g$ is going to provide small perturbations over the standard potential, let us consider a correction $c=10^{\,-1}$, so, the above equation implies that $\alpha \approx \pm\,24.61$, analogously, for a correction $c=10^{\,-2}$ we find $\alpha \approx \pm 204.96$. In such cases, if $k=\pm\,1$ we determine $g \approx 0.04$, and  $g \approx 0.005$ for $c=10^{\,-1}$, and $c=10^{\,-2}$, respectively. The effect of these perturbations caused by the Lorentz breaking parameter can be visualized in Figs. \ref{FIG2} and \ref{FIG3}, where it is shown that smaller values of $c$ indeed lead us to smaller corrections over the standard potential. Moreover, the fact that $k=(\alpha\,g)^{\,-1}$ means that the defects are sensible to the Lorentz-breaking parameter, as pointed in Fig. \ref{FIG1}, where we observe that the intensity of the transition between the two vacua depend on $g$.


\pagebreak

\subsection{Example - $\phi^{\,4}$ - $\chi^{\,4}$ with $r\neq0$}

As a matter of generalization, we apply the methodology tailored in this paper to the case  $\phi^{\,4}$ plus a $\chi^{\,4}$ model with $r\neq 0$. In this case, the superpotential $W$ has the form 
\be \label{ex_12}
W=k\,\left[(1-r^2)\, \phi + r\, \phi ^2-\frac{\phi ^3}{3}+\chi-\frac{\chi^{\,3}}{3}\right]\,.
\ee
So, substituting such superpotential back in Eq. \eqref{am_11} we find that functions $F$, and $G$ are
\be \label{ex_13}
F= k\,\left(1-(\phi -r)^2\right)+Z(\phi)\,(\phi-f(\chi))\,; \qquad G=k\,\left(1-\chi^{\,2}\right)+Z(\chi)\,(\phi-f(\chi))\,.
\ee

\begin{figure}[ht!]
\centering

\includegraphics[width=0.5\columnwidth]{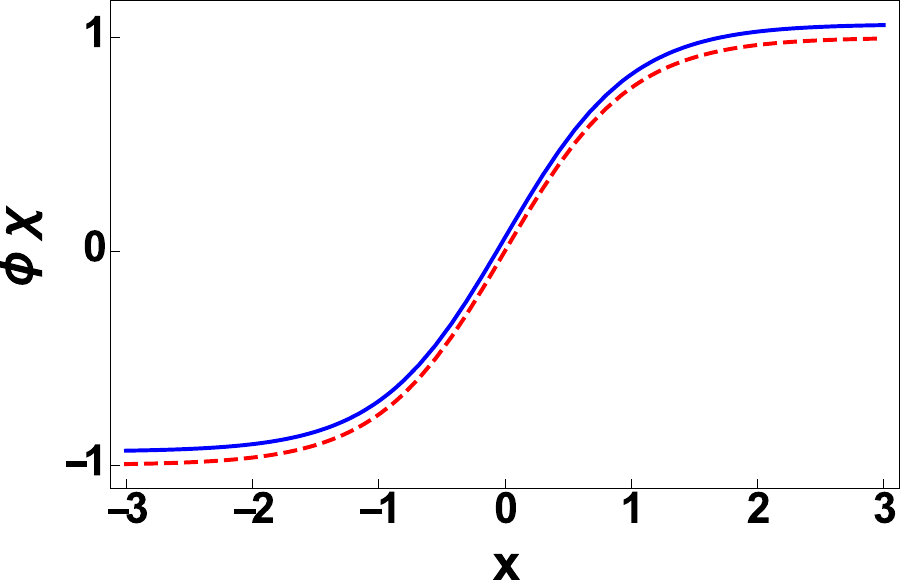}
\vspace{0.3cm}
\caption{Fields $\phi$ (solid blue curve), and $\chi$ (dashed red curve) for the $\phi^{\,4}$ - $\chi^{\,4}$ model with $r\neq 0$. The kinks were generated with $g=0.03$, and $r=1/16$.}
\label{FIG1r}
\end{figure}

Taking the same constraints for $Z$ pointed in Eq. \eqref{ex_4_2}, and supposing the {\it{Ansatz}}
\be \label{ex_14}
Z(\chi)= c_1\,\chi^{\,2}+c_2\,,
\ee
Eq. \eqref{am_10} is obeyed if
\be \label{ex_15}
c_1=\frac{8\, g\,k \,r+\sqrt{1-16\, g\,k\, r}-1}{8\, g\, r^2}\,;\qquad c_2=\frac{k-k\, \sqrt{1-16\, g\, k\, r}}{r\, \sqrt{1-16 g\, k\, r}+3\, r}\,,
\ee
and if we work with the straight line orbit
\be \label{ex_16}
\phi-\chi=r\,,
\ee
where $r$ is a dimensionless constant. {Once again, the straight line orbit imposes that $f_{\chi} =1 $, and $f_{\chi\,\chi} = 0$}. Therefore, we are able to write that
\be \label{ex_17}
Z(\phi)=c_1\,(\phi-r)^{\,2}+c_2\,,
\ee
resulting in the first-order differential equations
\ben \label{ex_18}
&&
\phi^{\,\prime}=k\,\left(1-(\phi-r)^{\,2}\right)+ \left(c_2+c_1 \,(\phi-r) ^2\right)\,(\phi-\chi)\,; \\ \nonumber
&&
\chi^{\,\prime}=k\,\left(1-\chi^{\,2}\right)+ \left(c_2+c_1\,\chi^{\,2}\right)\,(\phi-\chi)\,,
\een
whose analytical solutions are
\be \label{ex_19}
\phi(x)=A\,\tanh(\kappa\,x)+r\,; \qquad \chi(x)=A\,\tanh(\kappa\,x)\,,
\ee
where, 
\be \label{ex_20}
A^{\,2}=\frac{2\,k}{2\,k-c_1\, r}\,; \qquad \kappa^{\,2} =\frac{2\, k(k^{\,2}-c_1^{\,2}\, r)}{2\,k-c_1\,r}\,. 
\ee
The behavior of these kink-like defects are enhanced in Fig. \ref{FIG1r}, which reveals how the constant $r$ differs one defect from the other.

\begin{figure}[h!]

\includegraphics[width=0.45\columnwidth]{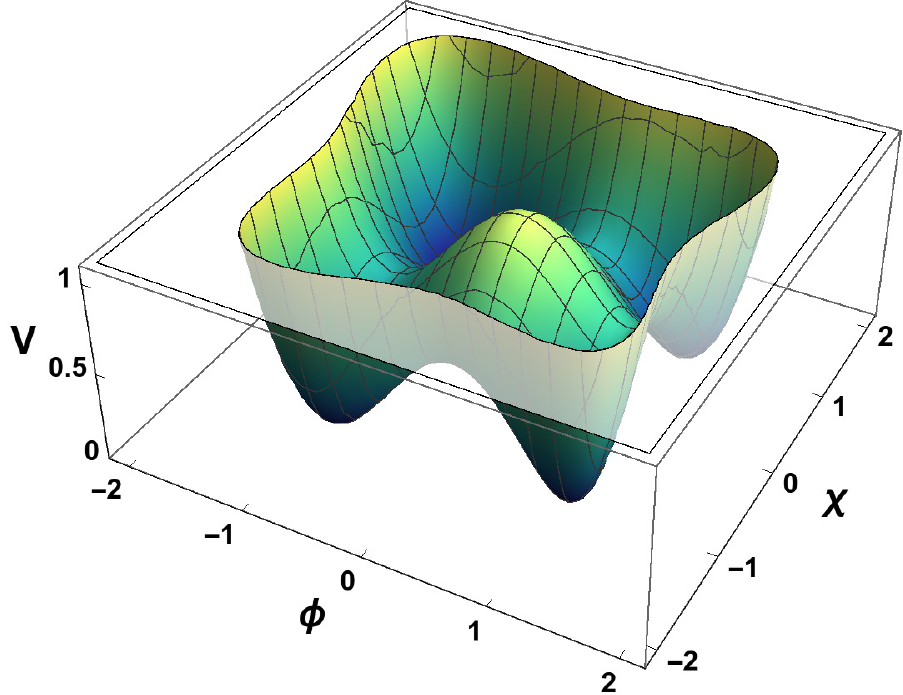} \hspace{0.2 cm} \includegraphics[width=0.45\columnwidth]{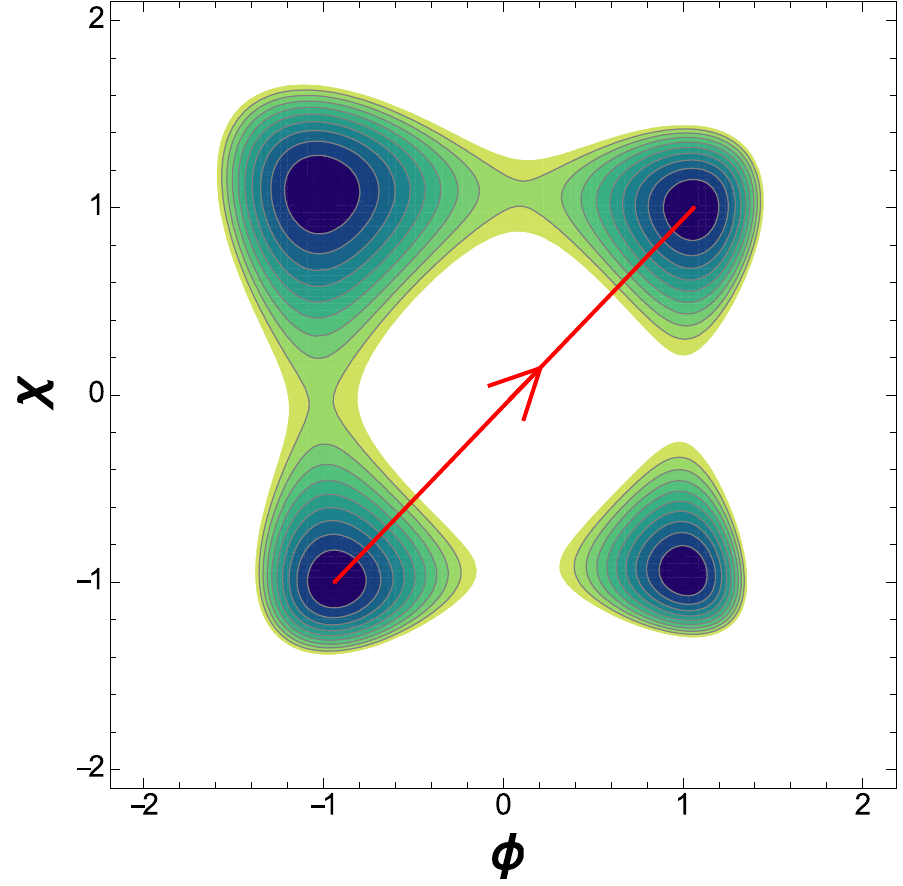}
\vspace{0.3cm}
\caption{The left panel shows potential $V$ Eq. \eqref{ex_21} for $g=0.03$, and $r=1/16$. The right graphic exhibits the contour of $V$, where the red line represents the analytical straight line orbit of our model. }
\label{FIG2r}
\end{figure}

\begin{figure}[h!]

\includegraphics[width=0.45\columnwidth]{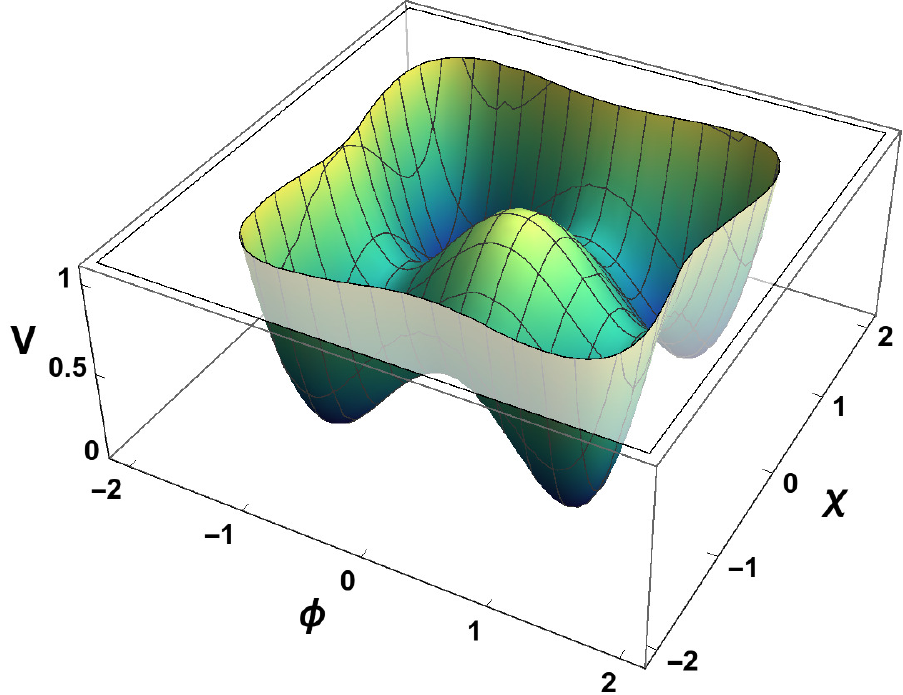} \hspace{0.2 cm} \includegraphics[width=0.45\columnwidth]{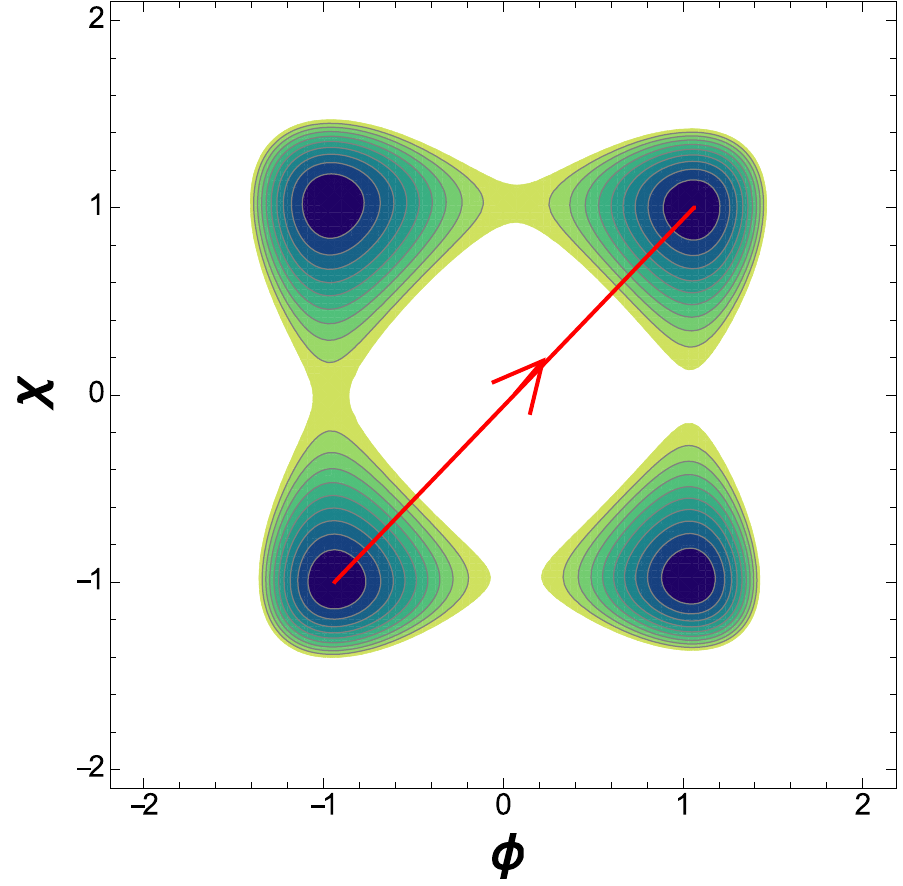}
\vspace{0.3cm}
\caption{The left panel shows potential $V$ Eq. \eqref{ex_21} for $g=0.01$, and $r=1/16$. The right graphic exhibits the contour of $V$, where the red line represents the analytical straight line orbit of our model. }
\label{FIG3r}
\end{figure}

From the previous first-order differential equations and from \eqref{am_4}, the effective two field potential $V(\phi,\chi)$ for this system is given by
\ben \label{ex_21} \nonumber 
&&
\hspace{-0.7 cm}V(\phi,\chi)= -g \bigg[\left(c_1 (\chi -\phi )+k\right) \left((\phi -\chi ) \left(c_1 (r-\phi )^2+c_2\right)+k \left(1-(r-\phi )^2\right)\right) \\ \nonumber
&&
\hspace{-0.7 cm}\times\,\left(\chi ^2 \left(3 c_2-c_1 (r-\phi )^2\right)+2 \chi ^3 (k-c_1 \phi )+3 c_1 \chi ^4-2 \chi  (c_2 \phi +k)-c_2 (r-\phi )^2\right) \\ \nonumber
&&
\hspace{-0.7 cm}-\left((\phi -\chi ) \left(c_1 \chi ^2+c_2\right)-k \chi ^2+k\right)\,\bigg(\left(\chi^{\,2}-(r-\phi )^2\right)\,\left(c_1 (r-\phi )^2+c_2\right) \\ \nonumber
&&
\hspace{-0.7 cm}\times\, (c_1 (\chi -\phi )+k)+2 k (r-\phi ) \bigg((\phi -\chi ) \left(c_1 (r-\phi )^2+c_2\right)+k \left(1-(r-\phi )^2\right)\bigg) \\ \nonumber
&&
\hspace{-0.7 cm}-2 c_1 (r-\phi ) (\phi -\chi ) \bigg((\phi -\chi ) \left(c_1 (r-\phi )^2+c_2\right)+k \left(1-(r-\phi )^2\right)\bigg)\bigg)\bigg] \\ \nonumber
&&
\hspace{-0.7 cm}+\frac{1}{2}\,\bigg[\left((\phi -\chi ) \left(c_1 (r-\phi )^2+c_2\right)+k \left(1-(r-\phi )^2\right)\right)^2 \\ 
&&
\hspace{-0.7 cm}+\left((\phi -\chi ) \left(c_1 \chi ^2+c_2\right)-k \chi ^2+k\right)^2\bigg]\,.
\een
The features of this potential can be appreciated in the graphics of Figs. \ref{FIG2r} and \ref{FIG3r}. There we observe that as the Lorentz parameter gets smaller, the deformation of potential $V$ decreases, corroborating with the previous example where $r=0$. Moreover, despite the increasing over the deformation, the orbit connected by fields $\phi$ and $\chi$ keeps unchanged. 

As in the previous example, we are now able to derive $V_\phi$, and $V_\chi$, revealing that the analytical solutions \eqref{ex_19} satisfy the equations of motion \eqref{am_2}. The Hamiltonian density related which such a model is
\ben \label{ex_22}
&& \nonumber
\hspace{-0.5cm}H=T^{\,0\,0}=-\frac{1}{2} \text{sech}^4(\kappa x) \bigg(\cosh (2 \kappa x) \left(\left(A^2-1\right) k-r \left(A^2\,c_1+c_2\right)\right)  \\ 
&&
+r\, \left(A^2\, c_1-c_2\right)-\left(A^2+1\right) k\bigg)^2\,(A\, g\, (k-c_1\, r) \tanh (\kappa x)+1)
\een
whose integration over the entire space results in
\be \label{ex_23}
E =  \frac{8}{3}\,k+\frac{8}{3}\,k^{\,2}\,r\, g+\frac{28}{3}\,k^{\,3}\,r^{\,2} \,g^2\,,
\ee
up to second order in $g$. 

It is relevant to point two features about this example, firstly that the dependences on the Lorentz breaking parameter $g$ of the solutions and the total energy, come directly when we deal with an orbit where $r\neq 0$. Secondly, that the total energy derived in Eq. \eqref{ex_23} is consistent with \eqref{ex_7_1} if we take $r=0$. Such a behavior unveils that $r$ can be thought as a correction for the orbit $\phi-\chi=0$, and that this correction is strongly sensible to the Lorentz breaking parameter. As we pointed before, the Lorentz breaking parameter is scaled by the Planck mass as follows
\be
g=\frac{\xi}{M_P}\,; \qquad M_P=L_P^{\,-1}\,,
\ee
where $L_p$ is the Planck length and $\xi$ is a dimensionless parameter.  Therefore, the products $\left(r\,g\right)^{\,n}$ are such that
\be
\left(r\,g\right)^{\,n}=r^{\,n}\,L_P^{\,n}\,\xi^{\,n}=r_P^{\,n}\,\xi^{\,n}\,; \qquad r_P=r\,L_P\,,
\ee
resulting in the energy
\be
E=\frac{8}{3}\,k+\frac{8}{3}\,k^{\,2}\,r_P\, \xi+\frac{28}{3}\,k^{\,3}\,r_P^{\,2} \,\xi^2\,.
\ee
This last approach informs that the orbit $r$ can be thought of as a dilatation of the Planck length, which induces the Lorentz violation effects into our topological defects. 


\pagebreak


\section{General analytical travelling wave solutions}
\label{atw}
Let us analyze the possibility of travelling wave solutions for timelike, spacelike and lightlike scenarios. In such an approach we are working with
\be \label{tw_1}
\phi=\phi(k\,x+\omega\,t)\,;\qquad \chi=\chi(k\,x+\omega\,t)\,,
\ee
which means that
\be \label{tw_2}
\dot{\phi}=\frac{\omega}{k}\,\phi^{\,\prime}\,; \qquad \ddot{\phi}=\frac{\omega^2}{k^2}\,\phi^{\,\prime\,\prime}\,;
\ee
\be\label{tw_3}
\dot{\chi}=\frac{\omega}{k}\,\chi^{\,\prime}\,; \qquad \ddot{\chi}=\frac{\omega^2}{k^2}\,\chi^{\,\prime\,\prime}\,.
\ee

Thus, Eqs. \eqref{tm01}, \eqref{sp1}, \eqref{ll1} can be rewritten as
\ben \label{tw_4}
&& 
-a\,\phi^{\,\prime\,\prime}+b_i\,\chi^{\,\prime\,\prime\,\prime}+V_{\,\phi}=0\,; \nonumber \\
&&
-a\,\chi^{\,\prime\,\prime}-b_i\,\phi^{\,\prime\,\prime\,\prime}+V_{\,\chi}=0\,,
\een
with $i=1$ for timelike, $i=2$ for spacelike and $i=3$ for lightlike, moreover, the explicit values of constants $a$ and $b_i$ are
\be \label{tw_5}
a=1-\frac{\omega^2}{k^2}\,;\,\, b_1=g\,\frac{\omega^2}{k^2}\,; \,\, b_2=g\,\frac{\omega}{k}\,;\,\, b_3=g\,\left(1+\frac{\omega}{k}\right)^{\,3}\,,
\ee
the last equation unveils that $a$ is nondimensional, while $b$ has dimension of $k^{\,-1}$.

A first integration of the equations of motion \eqref{tw_4} yields to the potential
\be \label{tw_5_1}
V(\phi,\chi)=a\,\left(\frac{\phi^{\,\prime\,2}}{2}+\frac{\chi^{\,\prime\,2}}{2}\right)-b_i\,\left(\chi^{\,\prime\,\prime}\,\phi^{\,\prime}-\phi^{\,\prime\,\prime}\,\chi^{\,\prime}\right)\,.
\ee
Therefore, considering the first-order differential equations established in Eq. \eqref{am_8}, and repeating the methodology presented in the last section, we find the equation of constraint
\ben \label{tw_5_2}
&&
a\,G_{\,\phi}+F_{\,\chi}\,\left(3\,b\,G_{\,\phi}-a\right)+b\,\left[G_{\,\chi}^{\,2}-G_{\,\chi}\,F_{\,\phi}+F_{\,\phi}^{\,2}\right. \\ \nonumber
&&
\left.+G\,\left(G_{\,\chi\chi}+F_{\,\phi\chi}\right)+F\,\left(G_{\,\phi\chi}+F_{\,\phi\,\phi}\right)\right]=0\,.
\een

One more time, the functions $F$, and $G$ obey the same prescriptions adopted in Eq. \eqref{am_11}.
\subsection{Example - BNRT model with $r=0$}

We exemplify the travelling wave scenario with a simple version of the so-called BNRT model introduced by Bazeia {\it et al.} \cite{bnrt}, whose superpotential is
\be\label{tw_6}
W=k\,\left(\phi-\frac{\phi^3}{3}\right)-k\,\phi\,\chi^{\,2}\,.
\ee
Again, for simplicity we take an orbit with $r=0$. This superpotential  allow us to write $F$, and $G$ in Eq. \eqref{am_11} as
\ben \label{tw_8}
&&
F=k \left(1- \phi ^2\right)-k\, \chi ^2+Z(\phi)\,\left(\phi -f(\chi)\right)\,; \\ \nonumber
&&
G=-2\, k\, \phi\,  \chi+Z(\chi)\left(\phi -f(\chi)\right)\,,
\een
so, substituting these functions in \eqref{am_10} we find
\ben \label{tw_8_1}
&&
-2 \,k\, \chi \, f\, \bigg\{f_{\chi}^2 \left(Z (a+2\, b\, k \,f+6\, b\, k\, \chi -2\, b\, Z)+4\, b\, k\, \chi\,  f\, Z_{\,\chi}\right) \\ \nonumber
&&
+f_\chi\, \left(Z (a-6\, b\, k\, \chi +b\, Z)+2\, b\, k\, f \left(Z \left(\chi\,  f_{\chi\,\chi}-1\right)+\chi \,Z_{\chi}\right)\right. \\ \nonumber
&&
\left.+b\, k\, f^2 \left(8\, k-Z_{\chi}\right)+4\, b\, k^2 \left(4\, \chi ^2-1\right)-b\, k\, \left(\chi ^2-1\right) \,Z_\chi\right) \\ \nonumber
&&
+b\, Z^{\,2} f_\chi^{\,3}-2 \,b\, k\, \left(f^2+\chi ^2-1\right)\, Z_{\chi}\bigg\}=0\,,
\een
where $Z=Z(\chi)$, and we used the prescriptions from \eqref{ex_4_2}. In this case, the orbit equation of $W$ which simplifies the expression above is
\be
\phi-\chi+1=0\,; \qquad f(\chi)=\chi-1\,; \qquad f_{\chi}=1\,; \qquad f_{\chi\,\chi}=0\,,
\ee
corresponding to a straight line orbit. Therefore, substituting the previous orbit in Eq. \eqref{tw_8_1}, we find
\be \label{tw_9}
Z(\chi)=-\frac{2}{a}\, b\, k^2\, \left(1-4\, \chi +6\, \chi ^2\right)\,.
\ee
Moreover, the orbit equation implies that 
\be \label{tw_10}
Z(\phi)=-\frac{2}{a}\, b\, k^2\, \left(3+8\, \phi +6\, \phi ^2\right)\,.
\ee

So, our first-order differential equations for this travelling wave model have the forms
\ben \label{tw_11}
\nonumber
&&
\phi^{\,\prime}=k \left(1- \phi ^2\right)-k\, \chi ^2  
-\frac{2}{a}\, b\, k^2\, \left(3+8\, \phi +6\, \phi ^2\right)\,\left(\phi -\chi +1\right)\,; \\
&&
\chi^{\,\prime}=-2\, k\, \phi\,  \chi-\frac{2}{a}\, b\, k^2\, \left(1-4\, \chi +6\, \chi ^2\right)\,\left(\phi -\chi +1\right)\,,
\een
whose analytical solutions are
\be \label{tw_7}
\phi(x,t)=-\frac{1}{4 e^{2\, k\,  x+2\,\omega\,t}+1}\,; \,\, \chi(x,t)=\frac{4 e^{2\, k\,  x+2\,\omega\,t}}{4 e^{2\, k\,  x+2\,\omega\,t}+1}\,.
\ee
The static versions of the solutions above, were introduced in \cite{dutraplb}.  In order to determine the effective potential, we substitute the differential equations from \eqref{tw_11}  into \eqref{tw_5_1}, leading to
\ben \label{tw_12}
\nonumber
&&
\hspace{-0.5 cm}V(\phi,\chi)=\frac{k^2}{2\, a^3}\, \bigg\{a^2\, \left[2\, b\, k\, \left(3+8\, \phi +6\, \phi ^2\right) (1+\phi -\chi )+a\, \left(\phi ^2+\chi ^2-1\right)\right]^2 \\ 
&&
\hspace{-0.5 cm}+4\, a^2\, \left[a \phi\,  \chi +b\, k\, (1+\phi -\chi ) \left(1-4\, \chi +6\, \chi ^2\right)\right]^2\\  \nonumber
&&
\hspace{-0.5 cm}-4\, b\, k\, \bigg[48\, b^3\, k^3 \left(3+8\, \phi +6\, \phi ^2\right) (1+\phi -\chi )^4\, \left(1-4\, \chi +6\, \chi ^2\right) \\ \nonumber
&&
\hspace{-0.5 cm}+a^2\, b\, k\, (1+\phi -\chi )^2 \bigg(2\, (\phi -1)\, \left(6\, \phi ^2-7\right)\, \chi
+\left(13+4\, \phi -30\, \phi ^2\right)\, \chi ^2 \\ \nonumber
&&
\hspace{-0.5 cm}-(\phi -1)^2-8\, (1+3\, \phi )\, \chi ^3-6\, \chi ^4\bigg)-a^3\, \chi\,  \left(\left(\phi ^2-1\right)^2-2\, \left(1+\phi ^2\right)\, \chi ^2+\chi ^4\right) \\ \nonumber
&&
\hspace{-0.5 cm} +4\, a\, b^2\, k^2\, (1+\phi -\chi )^3 \left(-5+2\, \phi +3\, \phi ^2+8\, (1+\phi ) \left(2-3\, \phi +9\, \phi ^2\right)\, \chi \right.\\ \nonumber
&& 
\hspace{-0.5 cm} \left.+(-53+6\, \phi  (2+9\, \phi ))\, \chi ^2+18\, \chi ^4\right)\bigg]\bigg\}\,.
\een

Thus, computing $V_\phi$, and $V_{\chi}$ it is possible to write the equations of motion \eqref{tw_4} explicitly, and also to check that the last are going to be satisfied by \eqref{tw_7}. Moreover, the Lagrangian density for the travelling wave approach is
\be \label{tw_13}
{\cal L}_{\,i}=-\frac{a}{2}\,\left(\phi^{\,\prime\,2}+\chi^{\,\prime\,2}\right)+b_{\,i}\,\chi^{\,\prime\,\prime}\,\phi^{\,\prime}-V(\phi,\chi)\,,
\ee
and following the procedure adopted in \eqref{ex_7_1} we see that $H_{i}=-{\cal L}_{i}$. Then, considering $V$ presented in Eq. \eqref{tw_12}, and the analytical solutions \eqref{tw_7} we are able to find
\ben \label{tw_14}
&&
H_i=\frac{128\, e^{4\,\xi}\, k^2\,\left(a - b_{\,i} \,k + 4\, e^{2 \,\xi}\, (a + b_{\,i}\, k)\right)}{\left(1 + 4\,e^{2\,\xi}\right)^5}\,; \\ \nonumber
&&
\xi=k\,x+\omega\,t\,,
\een
here $d\,\xi=k\,d\,x\,$, once the Hamiltonian density is invariant under travelling wave transformations. Thus, integrating $H_i$ in respect to variable $\xi$, we obtain the total energy $E$, which is given by
\be \label{tw_15}
E=\int_{-\infty}^{+\infty}\,d\,\xi\,\frac{H_i}{k}=\frac{2\, a\, k}{3}\,.
\ee
As in the previous section, we are able to trace a dependence between the Lorentz breaking parameter and $k$ (or $\omega$), analyzing the extreme points of our potential and comparing them with the standard form of the BNRT potential. Firstly, computing $V_{\,\phi\,\phi}$, and the determinant of the Hessian matrix $D=V_{\,\phi\,\phi}\,V{\,\chi\,\chi}-V_{\,\phi\,\chi}^{\,2}$ for points $(-1,0)$, and $(0,1)$ , we can see that they are truly minima of $V$ if
\be \label{tw_16}
2\,b_{\,i}\,k\,\left(1-\frac{3\,b_{\,i}\,k}{a}\right)<a\,,
\ee
leading to a first constraint involving $b$ and $a$. Moreover, the standard potential for the BNRT model which admits the orbits from \eqref{tw_7} is
\be \label{tw_17}
V_s(\phi,\chi)=a\,\frac{k^2}{2} \left(\left(\chi^2-1\right)^2+\phi^4+\left(6\, \chi ^2-2\right)\, \phi ^2\right)\,,
\ee
which presents four minima at $(\pm\,1,0)$, and $(0,\pm\,1)$, as well as a local maximum at $(0,0)$, where $V_s(0,0)=a\,k^{\,2}/2$. By substituting $\phi=\chi=0$ in potential $V(\phi,\chi)$ we have
\ben \label{tw_18}
&&
V(0,0)=\frac{k^2}{2}\, \bigg(a-12\, b_{\,i}\, k+\frac{44\, b_{\,i}^2\, k^2}{a}+\frac{80\, b_{\,i}^3\, k^3}{a^2}-\frac{576\, b_{\,i}^4\, k^4}{a^3}\bigg) \\ \nonumber
&&
\equiv V_{\,s}(0,0)-c\,a\,k^{\,2}=\left(\frac{1}{2}-c\right)\,a\,k^{\,2}\,,
\een
where $c$ is a correction to the standard potential caused by the Lorentz breaking parameter at $(0,0)$. Note that taking the limit $g\rightarrow 0$, and $c\rightarrow 0$ in the last equation, we determine
\be
V(0,0)=V_{\,s}(0,0)\,.
\ee
Then, once $b$ has dimension $k^{\,-1}$, repeating the procedure of the last section, we can define that $k\equiv a\,\left(\alpha\,b_{\,i}\right)^{\,-1}$, leading \eqref{tw_18} to
\be \label{tw_19}
\frac{\alpha\,\left[\alpha\,\left((\alpha -12) \alpha +44\right)+80\right]-576}{2\, \alpha ^4}=\frac{1}{2}-c\,,
\ee
so, if $c=10^{\,-1}$, we find $\alpha \approx 55.96$, and if $c=10^{\,-2}$, $\alpha \approx 596.30$. Furthermore, applying the definition $k \equiv a\,\left(\alpha\,b_{\,i}\right)^{\,-1}$ into \eqref{tw_5} results in the dispersion relations
\ben \label{tw_20}
&&
k_{\pm1} =\frac{\omega}{2} \,\left(\alpha\,  g\, \omega \pm\sqrt{4+\alpha ^2\, g^2\, \omega ^2}\right)\,; \\ \nonumber
&&
k_{\pm2} = \pm\,\frac{\omega }{\sqrt{1-\alpha\,  g\, \omega }}\,; \\ \nonumber
&&
k_{31}=-\omega\,; \qquad k_{\pm32}=\frac{1-2\, \alpha\,  g\, \omega \pm\,\sqrt{1-8\, \alpha\,  g\, \omega }}{2\, \alpha\,  g}\,, 
\een
therefore, when we take the limit $g\rightarrow 0$, we have $k_{\pm1}=k_{\pm2}=\pm\,\omega$, and $k_{31}=-\omega$, as it is required by a standard Lorentz invariant theory. Besides, expanding the dispersion relations for the lightlike scenario around small values of $g$, yields to
\be \label{tw_21}
k_{+32} =\omega+4\, \alpha\,\omega ^2\,g; \qquad k_{-32}= \frac{1}{\alpha\,  g}-3 \,\omega-4\,\alpha\,  \omega ^2\,g\,,
\ee
up to terms of order $g$. The last expression shows that $k_{-32}$ is strongly dependent on small values  of $g$, behaving like the dispersion relations found in section \ref{model}. Therefore, unlike $k_{31}$ and  $k_{+32}$,  $k_{-32}$ cannot reproduce the dispersion relation for a standard two-field Lagrangian density in a natural way, despite it leads to an integrable energy density. The nonanalytical behavior of $k_{-32}$ in the perturbative parameter is consistent with the dispersion relations obtained by Reyes in \cite{Reyes2,Reyes3} for the Myers and Pospelov  electromagnetic theory. As pointed in \cite{Reyes2,Reyes3} such a strong dependence on $g \rightarrow 0$ can be interpreted as extra degrees of freedom for the lightlike theory, characterizing it as a genuine higher-derivative theory. 

\begin{figure}[hb!]
\centering
\includegraphics[width=0.45\columnwidth]{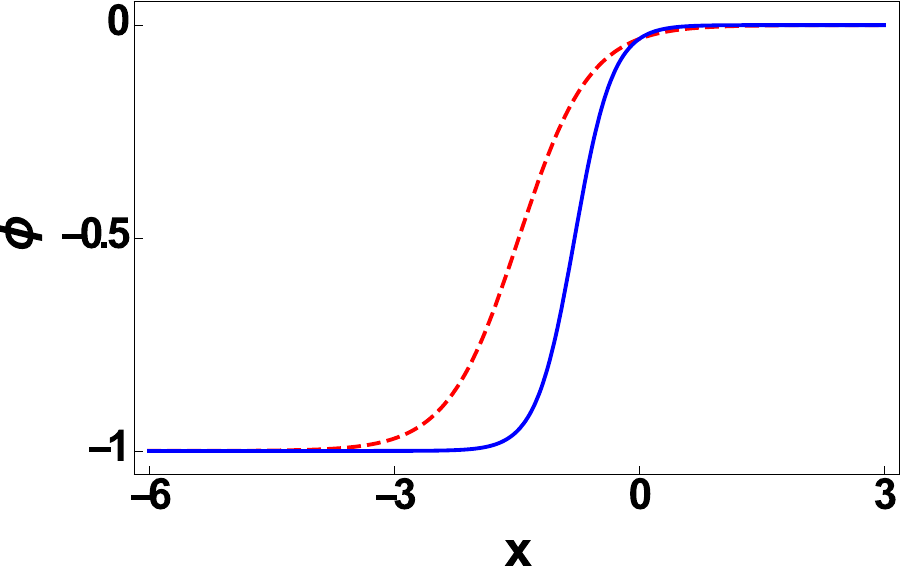} \hspace{0.1 cm} \includegraphics[width=0.45\columnwidth]{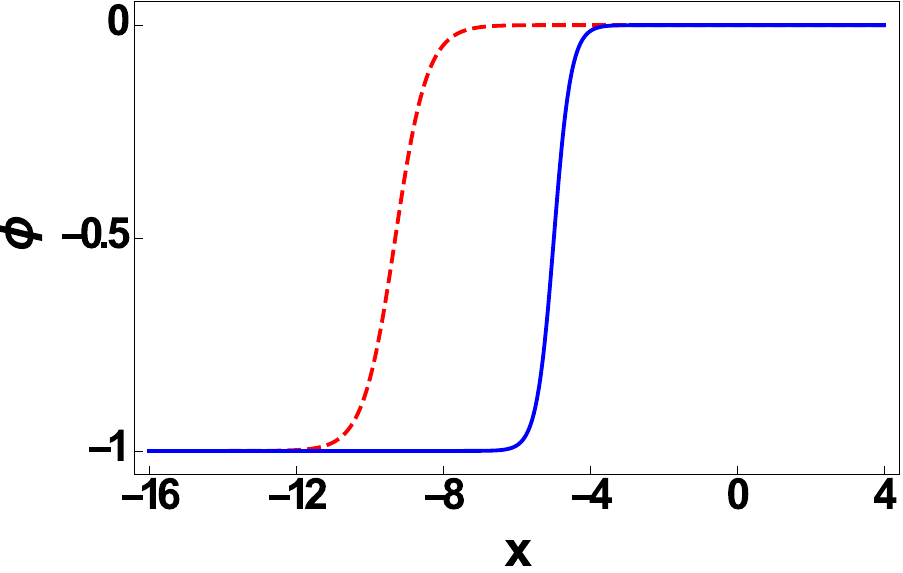}
\vspace{0.3cm}
\caption{ Field $\phi$ for the timelike BNRT model. The kinks were generated with $k_{+1}$ \eqref{tw_20}, $\alpha = 55.96$, $g=0.03$ (solid red curve), $g=0.005$ (dashed blue curve), $\omega=1$, $t=1$ (left panel), and $t=10$ (right panel). It is possible to realize that smaller values of $g$ result in faster displacements for the defects.}
\label{FIG4}
\end{figure}

\begin{figure}[ht!]
\vspace{0.3cm}
\includegraphics[width=0.45\columnwidth]{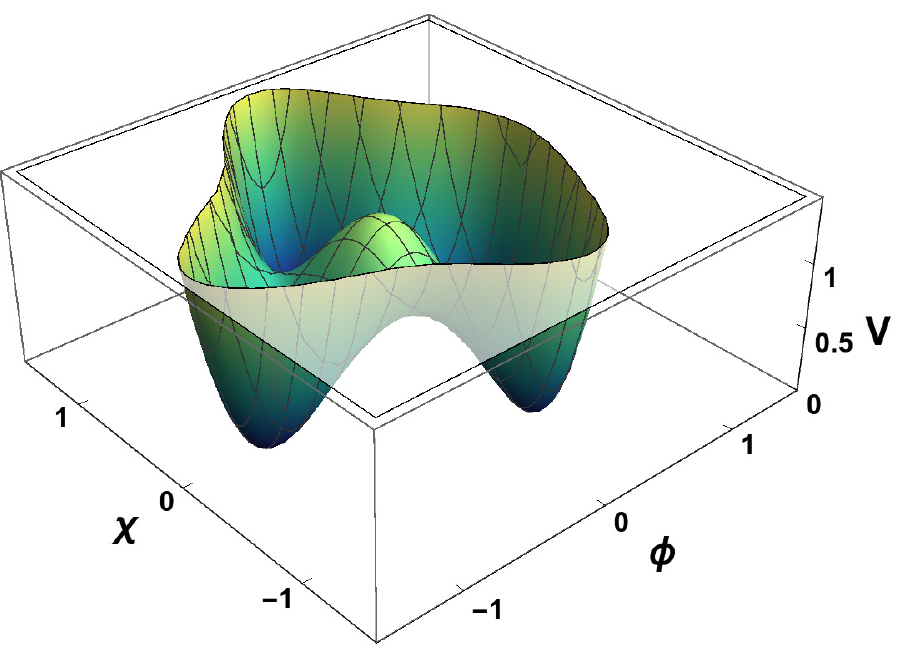} \hspace{0.1 cm} \includegraphics[width=0.45\columnwidth]{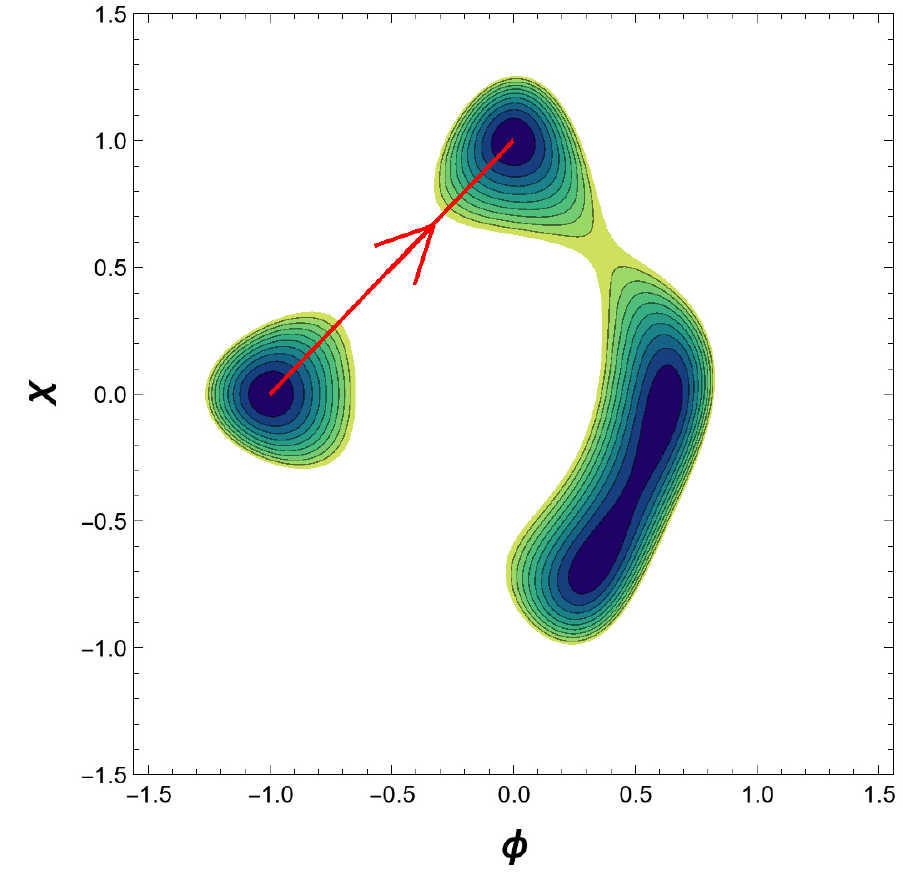}
\vspace{0.3cm}
\caption{The left panel shows  potential $V$ for the timelike scenario, with $k_{+1}=a\,\left(\alpha\,b_1\right)^{\,-1}$, $\alpha=55.96$, $k_{+1}=2$, $\omega=1$, and $g=0.03$. The right graphic exhibits the contour of $V$, where the red line represents the analytical straight line orbit of our model connecting the minimum $(-1,0)$ with $(0,1)$. }
\label{FIG5}
\end{figure}

\begin{figure}[hb!]
\vspace{0.3cm}
\includegraphics[width=0.45\columnwidth]{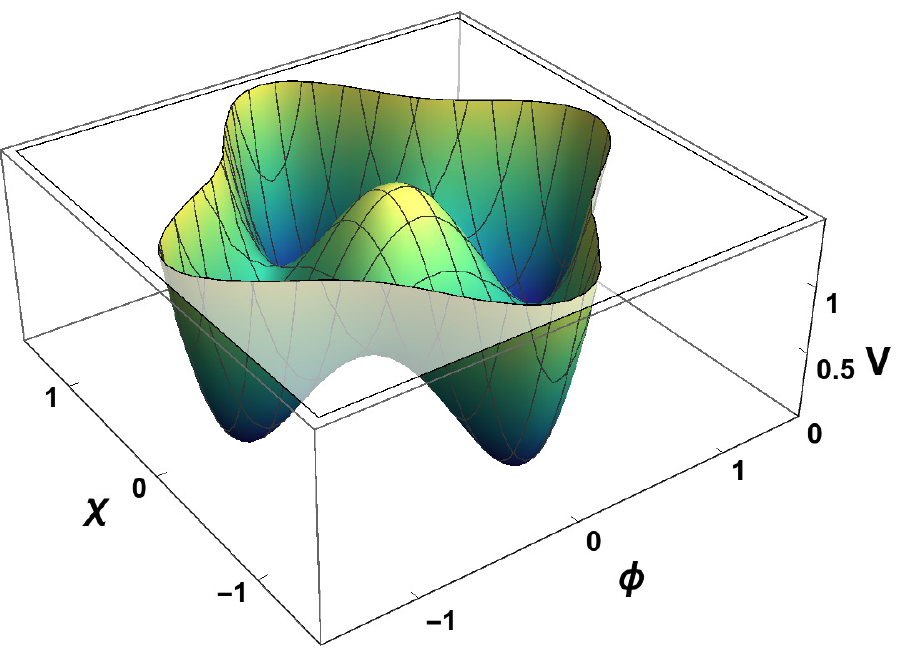} \hspace{0.1 cm} \includegraphics[width=0.45\columnwidth]{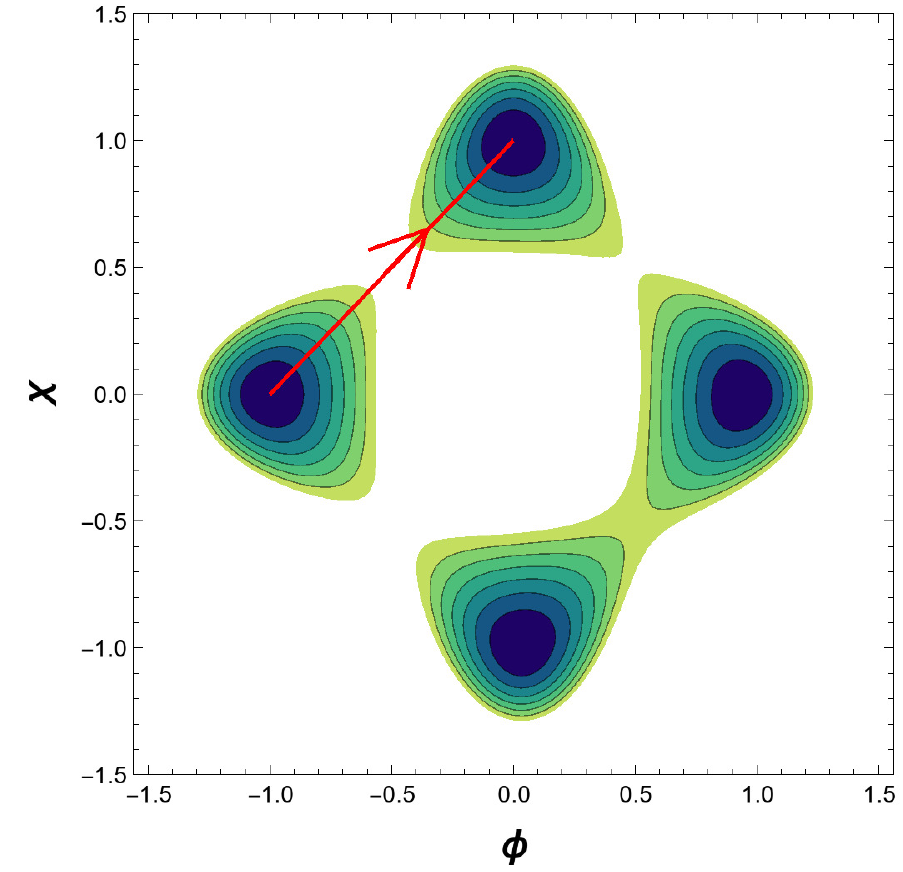}
\vspace{0.3cm}
\caption{The left panel shows  potential $V$ for the timelike scenario, with $k_{+1}=a\,\left(\alpha\,b_1\right)^{\,-1}$, $\alpha=596.30$, $k_{+1}=2$, $\omega=1$, and $g=0.003$. The right graphic exhibits the contour of $V$, where the red line represents the analytical straight line orbit of our model connecting the minimum $(-1,0)$ with $(0,1)$. }
\label{FIG6}
\end{figure}

In Fig. \ref{FIG4} we exemplify the travelling wave defects with the timelike scenario, there, one can see  that the defects move faster for smaller values of $g$. Besides, we plot the potential $V$ for the timelike scenario in Figs. \ref{FIG5}, and \ref{FIG6}, the first figure shows $V$ with a correction of $c=10^{\,-1}$ due the Lorentz breaking parameter, as well as its contour, where the minima $(-1,0)$, and $(0,1)$ are connect via the analytical solutions \eqref{tw_7}. The second figure unveils the behavior of $V$ with a correction $c=10^{\,-2}$ and its respective contour, where the red line shows our analytical orbits. The graphics allow us to visualize that smaller values of $c$ lead to smaller deformations in the potential of a Lorentz invariant theory. Another way to make the solitons depend on the breaking parameter, consists in add a perturbative term to the solutions of order $g$, as we are going to show below.


\section{Perturbative Approach}
\label{pertapp}

In this approach we are going to follow the procedure adopted by \cite{bazeia2010lv}, to verify what is the influence of small perturbations of order $g$ over the analytical defects. In order to see the features of this effect, let us consider the following definitions:
\be \label{pt_1}
\phi \rightarrow \phi + \epsilon\,\zeta \,; \qquad \chi \rightarrow \chi +\epsilon\,\rho\,
\ee
where $\zeta$ and  $\rho$ are also travelling wave solutions, which means $\zeta=\zeta(k\,x+\omega\,t)$ and $\rho=\rho(k\,x+\omega\,t)$. 




The previous definitions make the potential $V(\phi,\chi)$ be rewritten as
\be \label{pt_2}
V(\phi,\chi) \rightarrow V(\phi,\chi)+\epsilon\,V_{\phi}\,\zeta+\epsilon\,V_{\chi}\,\rho\,,
\ee
up to first order terms in $\epsilon$. Thus, substituting these ingredients into the equations of motion for the three different scenarios we find 
\ben \label{pt_3}
&&
-\epsilon\,a\,\zeta^{\,\prime\,\prime}+\epsilon\,b_i\,\rho^{\,\prime\prime\prime}+\epsilon\,V_{\,\phi\,\phi}\,\zeta+\epsilon\,V_{\,\chi\,\phi}\,\rho=0\,; \\ \nonumber
&&
-\epsilon\,a\,\rho^{\,\prime\,\prime}-\epsilon\,b_i\zeta^{\,\prime\prime\prime}+\epsilon\,V_{\,\chi\,\phi}\,\zeta+\epsilon\,V_{\,\chi\,\chi}\,\rho=0\,,
\een
plus the unperturbed equations of motion. From Eqs. \eqref{tw_5}, \eqref{tw_9}, and \eqref{tw_10} we see that $b_i \propto g$, and $Z \propto g$, therefore \eqref{pt_3} is going to have terms of order $\epsilon$ and $\epsilon\,g$. The terms of order $\epsilon\,g$ can be ignored if $\epsilon \equiv \beta\,g$, where $\beta$ is a constant with dimension $g^{\,-1}$. Such a procedure also reduces the number of arbitrary parameters. Then, considering $\epsilon=\beta\,g$ in \eqref{tw_5_1} the derivatives of $V$ are
\be \label{pt_4}
\epsilon\, V_{\,\phi\phi} = \beta\,g\,a\,\left(W_{\,\phi\phi\phi}\,W_\phi+W_{\,\phi\phi}^{\,2} +W_{\,\phi \chi}^{\,2}+W_{\,\chi}\,W_{\,\phi\phi\chi}\right)\,;
\ee
\be \label{pt_5}
\epsilon\,V_{\,\chi\chi} = \beta\,g\,a\left(W_{\,\chi\chi\chi}\,W_\chi+W_{\,\chi\chi}^{\,2}+W_{\,\phi \chi}^{\,2}+W_{\phi}W_{\,\phi\chi\chi}\right)\,;
\ee
\ben \label{pt_6}
\epsilon\,V_{\,\chi\phi} &=& \beta\,g\,a\,\big(W_{\,\chi\chi\phi}\,W_\chi+W_{\,\chi\chi}W_{\,\chi\phi}\\ \nonumber
&&
+W_{\,\phi\phi}\,W_{\,\phi \chi}+W_{\,\phi}W_{\,\phi\phi\chi}\big)\,,
\een
where we eliminated terms of order $g^2$.

If the superpotential does not present interacting terms involving the fields, which means $W=\widehat{W}(\phi)+\widetilde{W}(\chi)$, then the effective equations for $\zeta$ and $\rho$ can be easily decoupled to
\ben \label{pt_7}
&&
\zeta^{\,\prime\,\prime}=\left(\widehat{W}_{\,\phi\,\phi\,\phi}\,\widehat{W}_{\,\phi}+\widehat{W}_{\,\phi\,\phi}^{\,2}\right)\,\zeta\,; \\ \nonumber
&&
\rho^{\,\prime\,\prime}=\left(\widetilde{W}_{\,\chi\,\chi\,\chi}\,\widetilde{W}_{\,\chi}+\widetilde{W}_{\,\chi\,\chi}^2\right)\,\rho\,.
\een
Let us exemplify this methodology with a $\phi^{\,4}$ - $\chi^{\,4}$ superpotential, then substituting $W$ from Eq. \eqref{ex_1} together with the solutions
\be \label{pt_8}
\phi(x,t)=\chi(x,t)=\tanh(k\,x+\omega\,t)\,,
\ee
into \eqref{pt_7}, we find the second-order differential equations
\ben \label{pt_9}
&&
\zeta^{\,\prime\,\prime}=2\,k^{\,2}\,\left(3\,\tanh^{\,2}(k\,x+\omega\,t)-1\right)\,\zeta;\\ \nonumber
&&
\rho^{\,\prime\,\prime}=2\,k^{\,2}\,\left(3\,\tanh^{\,2}(k\,x+\omega\,t)-1\right)\,\rho\,,
\een
whose analytical solutions are
\be
\zeta=\rho=\mbox{sech}^{\,2}\,(k\,x+\omega\,t)\,,
\ee
therefore, our defects, up to small perturbations on $g$, have the forms
\be
\phi=\chi=\tanh(k\,x+\omega\,t)+\beta\,g\,\mbox{sech}^{\,2}\,(k\,x+\omega\,t)\,.
\ee

The last result unveil that the soliton solutions are sensible to perturbations caused by the symmetry breaking process. Furthermore, we can also observe that the topological sector connected by the defects stills unchanged at least up to approximations of order $g$.

\section{Final Remarks}\label{remarks}
In this work we studied carefully the 2D Myers-Pospelov Lagrangian, which was written in terms of two real scalar fields $\phi$ and $\chi$. In our investigations, we were able to derive the equations of motion for three distinct scenarios denominated by timelike, spacelike, and lightlike. Our first analyses concerning the equations of motion of these different approaches showed that they are satisfied for a specific set of analytical fields which are solutions of the Burgers equation at low velocities. We introduced a procedure to obtain analytical solutions for static, as well as for travelling wave fields related with such a Lagrangian. 
In order to have a better perception about the applicability of the procedure, we exemplified it with static and time-dependent defect-like fields. We applied the procedure for the a static $\phi^{4}$ - $\chi^{\,4}$ superpotential, and for the BNRT model, which is the simplest solvable example of a two scalar coupled fields model. Our methodology resulted in potentials which have several coupling terms involving both fields. As we pointed in Figs \ref{FIG2}, \ref{FIG3}, \ref{FIG2r}, \ref{FIG3r}, \ref{FIG5}, and \ref{FIG6} the Lorentz breaking parameter is responsible for the deformation effects exhibited in our potentials. In orbits where $r=0$, the dependence between $k$ and $g$ in sections \ref{asm}, and \ref{atw} were defined in such a way that the derived analytical models are consistent with small deformations over the standard potentials. Moreover, $g$ parameter also can change the intensity of the transition between two vacua, as observed in Fig. \ref{FIG1} or the velocity of the propagation of the soliton, vide Fig. \ref{FIG4}. Furthermore, we built an analytical static model for an orbit with $r\neq0$, which resulted in another example of implement a dependence between $k$ and $g$. This last approach reveled that the effects derived from $r$ in the total energy of this model is extremely sensible to the Lorentz breaking parameter. We also considered another alternative to include the Lorentz breaking terms into the soliton solutions via a perturbative approach. We exemplified this approach with a $\phi^{4}$ - $\chi^{\,4}$ travelling wave model, leading us to analytical differential equations for the perturbed waves $\zeta$, and $\rho$. An interesting fact is that the perturbed solutions do not change the vacua of the original fields, at least for perturbations of order $g$. It is remarkable that the methodology tailored in this study leaded to analytical models, despite the complexity of the equations of motion derived from the 2D Myers-Pospelov Lagrangian. We believe that the introduced methodology unveil a new route to determine analytical solutions for models with generalized kinematic terms, and it can be easily applied to other scenarios like in the famous Podolsky electromagnetism \cite{podolsky}. Investigations in this direction are in process and we hope to report on it in near future.

  
\section*{Acknowledgements}
We would like to thank CNPq, CAPES, PNPD/PROCAD-CAPES for financial support. E. Passos also would like to thank the hospitality of the Instituto de Fı\'isica at the Universidade Federal do Rio de Janeiro, where part of this work was constructed.

\end{document}